\documentclass[aps,prd,preprint,groupedaddress,amstex]{revtex4}

\input axodraw.sty

\def\be{\begin{equation}}
\def\ee{\end{equation}}
\def\beq{\begin{eqnarray}}
\def\eeq{\end{eqnarray}}
\def\s{\sigma}

\def\G{\Gamma}

\def\an{analytic}
\def\ac{\an{} continuation}

\def\hf{hypergeometric function}

\def\lra{\leftrightarrow}
\def\ai{\alpha_1}
\def\aii{\alpha_2}
\def\aiii{\alpha_3}
\def\aiv{\alpha_4}
\def\av{\alpha_5}
\def\avi{\alpha_6}

\begin{document}


\title{General Formula for N-point One-loop Feynman Integrals}


\author{Alfredo T. Suzuki, Esdras S. Santos}
\email[]{suzuki, esdras@ift.unesp.br}
\affiliation{Instituto de F\'{\i}sica Te\'orica - UNESP, \\
R.Pamplona, 145 S\~ao Paulo SP, CEP 01405-900 Brazil}


\author{Alexandre G. M. Schmidt}   
\email[]{schmidt@fisica.ufpr.br}
\affiliation{Departamento de F\'{\i}sica --- Universidade Federal do Paran\'a \\
Caixa Postal 19044, Curitiba PR, CEP 81531-990 Brazil}

\date{\today}

\begin{abstract}
The negative dimensional integration method (NDIM) is a technique
where several difficulties concerning loop integration can be
overcome. From usual covariant gauges to complicated Coulomb gauge
integrals, and even the trickiest light-cone integrals one can
apply the methodology of NDIM. In this work we show how to
construct a general formula --- we mean arbitrary exponents of
propagators, off-shell external momenta and distinct massive
propagators --- for one-loop scalar integrals, for {\it covariant}
gauges, and apply it to one through six-point loop integrals. We
present detailed analysis of pentagon and hexagon scalar integrals
for massive/massless internal particles being external momenta on
or off mass shell.
\end{abstract}

\pacs{02.90+p, 12.38.Bx}
\keywords{radiative corrections, n-point one-loop integrals,
negative-dimensional integration}

\maketitle


\section{Introduction}

Important mathematical methods have been required to evaluation of
the complex Feynman integrals in the calculations of scattering
amplitudes in QED and QCD, in the radiative corrections, study of
Green function behavior, renormalization group and others problems
in quantum field theory. The integration using Mellin-Barnes
representation \cite{davyd,davyd2,boos}, the Gegenbauer polynomial
technique \cite{gegenbauer}, integration by parts \cite{tkachov},
negative dimensional integration (NDIM) \cite{halliday,box},
string inspired methods\cite{string}, differential equation
approach\cite{gehrmann} and several others
\cite{smirnov,laporta,anastasiou,bern,chetyrkin,glover,tausk,fleischer,malbouisson,lozano,bashir,bern-gravit},
are some of the technique that have been currently used.

In the present article we make use of the NDIM
\cite{halliday,box}. Its implementation transfer the complexities
of the performing $D$-dimensional Feynman integrals to resolution
of a system of linear algebraic equations that we call of system
of constraint equations. Therefore, when we choose one solution to
those constraint equations we are obtaining one specific solution
to Feynman integral in question that is defined in a specific
kinematic region. The others solutions to system constraint
equations represent the possibilities of analytic continuation of
such solution. In other words, the NDIM solve the Feynman integral
as well as obtain solving the constraint system equation, all the
analytic continuation possibilities to the solution found. We note
that after arrive the final formula to Feynman integral is
necessary yet carry out other but straightforward analytic
continuation, now to negative value of the powers of the
propagators in the scalar integral.

We implement the NDIM in this paper to obtain the one-loop general
massive n-point function. As already mentioned above we obtain, by
solving the system constraints equation, all the other solutions
analytically continued and show that the number of this
possibilities is a function of number of external lines $n$. We
 present then applications to cases of 1 to 6-point functions, considering only
 scalar integrals since any tensorial integral
can be reduced to scalar integral according to \cite{passarino}.
The hypergeometric series we choose to present in this work are
such that can be used in the dimensional or analytic
regularization schemes \cite{giambiagi,leibbrandt,milgram}, always
preserving gauge symmetry, a fact well-known from quantum field
theory. Similar results for N-point scalar integrals were obtained
in \cite{davyd,davyd2} using Mellin-Barnes approach, however
Davydychev quote only the hypergeometric series he call
"symmetric". Here we write such results and several others, an
interesting feature of NDIM, providing a very large number of
hypergeometric series that represent the original Feynman
integral.

This work is organized as follows. In the early section we present
a detailed approach to the implementation of the NDIM to one-loop
n-point function. In the subsequent sections we have applied the
results from section one, the general formula, starting from
one-point integrals, two, three, four (box), pentagon and finally
hexagon scalar integrals. These results are exact, i.e., no
approximations was made. The solutions obtained in this sections
are given in terms of the hypergeometric functions (see appendix)
and compared with known ones, when they are available, in the
literature.

\section{One-loop n-point function: general formula}

In this section we present the calculations to evaluate the
one-loop scalar integrals with the NDIM. Consider a one-loop
Feynman diagram  with $n$, $n=1,2,...,$ internal momenta
$l_0,l_0-l_1,l_0-l_2,...,l_0-l_{n-1}$ and masses
$m_0,m_1,m_2,...,m_{n-1}$, where $l_1,l_2,...,l_{n-1}$ are given
in terms of a linear combination of the external momenta. Its
scalar integral associated is
\begin{equation}
\int \frac{d^{D}l_{0}}{%
[l_{0}^{2}-m_{0}^{2}]^{a_{0}}[(l_{0}-l_{1})^{2}-m_{1}^{2}]^{a_{1}}...[(l_0-l_{n-1})^{2}-m_{n-1}^{2}]^{a_{n-1}}%
}.  \label{a1}
\end{equation}%
Consider now the gaussian integral
\begin{equation}
I=\int d^{D}l_{0}\exp \{-\alpha
_{0}[l_{0}^{2}-m_{0}^{2}]-{\sum_{i=1}^{n-1}}\alpha
_{i}[(l_{0}-l_{i})^{2}-m_{i}^{2}]\}.  \label{a2}
\end{equation}%
where $\alpha _{0}$, $\alpha _{i}$ are positive parameters. Then,
the exponential function above can be expanded and we get
\begin{eqnarray}
I &=&{\sum_{a_{0},...,a_{n-1}=0}^{\infty }}(-1)^{ {\sum
_{i=0}^{n-1}}a_{i}}\frac{\alpha _{0}^{a_{0}}\alpha
_{1}^{a_{1}}....\alpha
_{n-1}^{a_{n-1}}}{a_{0}!a_{1}!...a_{n-1}!}  \nonumber \\
&&\times \int
d^{D}l_{0}[l_{0}^{2}-m_{0}^{2}]^{a_{0}}[(l_{0}-l_{1})^{2}-m_{1}^{2}]^{a_{1}}...[
(l_{0}-l_{n-1})^{2}-m_{n-1}^{2}]^{a_{n-1}}. \label{a3}
\end{eqnarray}

Using the definition
\begin{eqnarray}
\alpha  &=&\alpha _{0}+{\sum_{i=1}^{n-1}}\alpha _{i}, \label{a4}
\end{eqnarray}
we can rewrite the $I$ integral of form
\begin{eqnarray}
I &=&\int d^Dl_{0}\exp \{-\alpha[l_0^2-2\frac{ \sum_{i=1}^{n-1}
\alpha_{i}l_{0}\cdot l_{i}}{\alpha}+\frac{(\sum_{i=1}^{n-1}\alpha_{i}l_{i})^2}{\alpha^2}] \nonumber \\
&&+\frac{\left(\sum_{i=1}^{n-1}\alpha_il_i\right)^2}{\alpha}-\sum_{i=1}^{n-1}\alpha_{i}l_{i}^2+\sum_{i=0}^{n-1}\alpha_{i}m_{i}^2\}.
\end{eqnarray}
After evaluate the integral we have
\begin{eqnarray}
I &=&(\frac{\pi}{\alpha})^{D/2}\exp \{\frac{\left(
\sum_{i=1}^{n-1}\alpha_il_i\right)^2}{\alpha}-\sum_{i=1}^{n-1}\alpha_{i}l_{i}^2+\sum_{i=0}^{n-1}\alpha_{i}m_{i}^2\}.
\end{eqnarray}
Yet can us rewritten this expression of form
\begin{eqnarray}
I &=&(\frac{\pi}{\alpha})^{D/2}\exp \{\frac{
\sum_{i=1}^{n-1}\alpha_{i}\alpha_{j}l_{i}\cdot l_{j}-\alpha_{0}
{\sum_{i=1}^{n-1}}\alpha_{i}l_{i}^{2}-
{\sum_{i=1}^{n-1}}\alpha_{i}\alpha_{j}l_{i}^2}{\alpha} \nonumber \\
&&+\sum_{i=0}^{n-1}\alpha_{i}m_{i}^2 \}. \label{a5}
\end{eqnarray}
Also can be show that
\begin{eqnarray}
\sum_{i,j=1}^{n-1}\alpha_{i}\alpha_{j}l_{i}\cdot l_{j}&=&2\sum_{i>j=1}^{n-1}\alpha_{i}\alpha_{j}l_{i}\cdot l_{j}+\sum_{i=1}^{n-1}\alpha_{i}^{2}l^2_{i},\\
\sum_{i,j=1}^{n-1}\alpha_{i}\alpha_{j}l_{i}^2&=&\sum_{i>j=1}^{n-1}\alpha_{i}
\alpha_{j}(l_{i}^2+l_{j}^2)+\sum_{i=1}^{n-1}\alpha_{i}^{2}l_{i}^{2}.
\end{eqnarray}
Performing the substitution of this results in (\ref{a5}), we get
\begin{eqnarray}
I &=&(\frac{\pi}{\alpha})^{D/2}\exp \{\frac{-
{{\sum_{i>j=1}^{n-1}}}\alpha_{i}\alpha_{j}l_{ij}^2-\alpha_{0}
{{\sum_{i=1}^{n-1}}}\alpha_{i}l_{i}^{2}}{\alpha} \nonumber \\
&&+\sum_{i=0}^{n-1}\alpha_{i}m_{i}^2, \}
\end{eqnarray}
where $l_{ij}=l_{i}-l_{j}$. From exponential argument above we
have  $w=\frac{n^2-3n+2}{2}$ terms with different coefficients
$\frac{\alpha_{i}\alpha_{j}}{\alpha}$, namely
\[
\begin{tabular}{ccccc}
$\alpha _{1}\alpha _{2},$ & $\alpha _{1}\alpha _{3},$ & $\alpha
_{1}\alpha
_{4},$ & $...$ & $\alpha _{1}\alpha _{n-1},$ \\
& $\alpha _{2}\alpha _{3},$ & $\alpha _{2}\alpha _{4},$ & $...$ &
$\alpha
_{2}\alpha _{n-1},$ \\
&  & $\alpha _{3}\alpha _{4},$ & $...$ & $\alpha _{3}\alpha _{n-1},$ \\
&  &  & $...$ &  \\
&  &  & $...$ & $\alpha _{n-2}\alpha _{n-1}.$%
\end{tabular}%
\
\]
Also, there are $n-1$ terms with different coefficients
$\alpha_{0}\alpha_{i}$ and $n$ terms to coefficients $\alpha_{i}$.
This result in $w+2n-1=\frac{n^2+n}{2}$ terms with different
coefficients. After expansion of the exponential above We can
write
\begin{eqnarray}
I &=&(\frac{\pi}{\alpha})^{D/2}\sum_{j_{1},...,j_{w+2n-1}}^{\infty}\frac{1}{\alpha^{j_{1}+j_{2}+...+j_{w+n-1}}}\nonumber \\
&&\times \alpha_{0}^{j_1+j_2+...+j_{n-1}}\alpha_{1}^{j_1+j_n+...+j_{2n-2}}\nonumber \\
&&\times \alpha_{2}^{j_2+j_n+j_{2n-2}+...+j_{3n-4}}... \alpha_{n-1}^{j_{2n-2}+j_{3n-4}+j_{4n-7}+...+j_w} \nonumber \\
&&\times \frac{(-l_1^2)^{j_1}}{{j_1}!}\frac{(-l_2^2)^{j_2}}{{j_2}!}...\frac{(-l_{n-1}^2)^{j_{n-1}}}{{j_{n-1}}!} \nonumber \\
&&\times
\frac{(-l_{12}^2)^{j_n}}{{j_n}!}\frac{(-l_{13}^2)^{j_{n+1}}}{{j_{n+1}!}}
...\frac{(-l_{n-1,n-2}^2)^{j_{w+n-1}}}{{j_{w+n-1}}!} \nonumber \\
&&\times
\frac{(m_{0}^2)^{j_{w+n}}}{{j_{w+n}}!}\frac{(m_{1}^2)^{j_{w+n+1}}}{{j_{w+n+1}}!}...\frac{(m_{n-1}^2)^{j_{w+2n-1}}}{j_{w+2n-1}!}.
\label{a4}
\end{eqnarray}
If we take the multinomial expansion in the exponents of the
$\alpha$, given by (\ref{a4}), we get
\begin{eqnarray}
\frac{1}{\alpha ^{D/2+j_{1}+j_{2}+...+j_{w+n-1}}} &=&\frac{1}{[\alpha _{0}+{%
{{\sum_{i=1}^{n-1}}}}\alpha _{i}]^{D/2+j_{1}+j_{2}+...+j_{w+n-1}}}
\nonumber \\
&=&\sum_{j_{w+2n},...,j_{w+3n-1}}^{\infty}\Gamma(1-D/2-j_{1}-j_{2}-...-j_{w+n-1})  \nonumber \\
&&\times \frac{\alpha _{0}^{j_{w+2n}}}{j_{w+2n}!}\frac{\alpha
_{1}^{j_{w+2n+1}}}{j_{w+2n+1}!}...\frac{\alpha
_{n-1}^{j_{w+3n-1}}}{j_{w+3n-1}!},
\end{eqnarray}%
with the constraint
\begin{equation}
D/2=-j_{1}-j_{2}-...-j_{w+n-1}-j_{w+2n}-j_{w+2n+1}-...-j_{w+3n-1}.
\nonumber
\end{equation}%
Performing the substitution of this expression in (\ref{a4}) and
compare the exponents of the parameters $\alpha_{i}$ in
(\ref{a3}), we have follow constraint equations
\begin{eqnarray}
a_{0} &=&j_{1}+j_{2}+...+j_{n-1}+j_{w+n}+j_{w+2n}, \\
a_{1} &=&j_{1}+j_{n}+...+j_{2n-2}+j_{w+n+1}+j_{w+2n+1}, \\
a_{2} &=&j_{2}+j_{n}+j_{2n-2}+...+j_{3n-6}+j_{w+n+2}+j_{w+2n+2}, \\
a_{3} &=&j_{3}+j_{n+1}+j_{2n-2}+j_{3n-5}+...+j_{4n-10}+j_{w+n+3}+j_{w+2n+3}\\
.  \nonumber \\
.  \nonumber \\
.  \nonumber \\
a_{n-1} &=&j_{n-1}+j_{2n-2}+j_{3n-6}+j_{4n-10}+...\nonumber \\
&&...+j_{w+n-1}+j_{w+2n-1}+j_{w+3n-1}\\
D/2
&=&-j_{1}-j_{2}-...-j_{w+n-1}-j_{w+2n}-j_{w+2n+1}-...-j_{w+3n-1}.
\end{eqnarray}%
Using the results we have $S=w+3n-1=\frac{n^{2}+3n}{2}$ sums with
$n+1$ constraint equations and $C_{S,n+1}$ different forms to
evaluate. But, if there are $F_{i}$ internal lines or $F_{e}$
external lines  associated to no massive fields, the number of the
sums reduce to $S-F_{i}-F_{e}$. Then, the number of the different
forms to perform the  $S-F_{i}-F_{e}$ sums is
\begin{equation}
C_{S-F_{i}-F_{e},n+1}=\frac{(\frac{n^{2}+3n}{2}-F_{i}-F_{e})!}{(n+1)!(%
\frac{n^{2}+n-2}{2}-F_{i}-F_{e})!}.
\end{equation}%

Finally we obtain the final expression to integral given in
(\ref{a3}), that is
\begin{eqnarray}
J^{(n)}&=&J^{(n)}{(D,a_0,...,a_{n-1},l_1,l_2,...,l_{n-1},m_0,m_1,...m_{n-1})}\nonumber  \\
&=&\int d^{D}l_{0}[l_{0}^{2}-m_{0}^{2}]^{a_{0}}[(l_{0}-l_{1})^{2}-m_{1}^{2}]^{a_{1}}...[(l_{0}-l_{n-1})^{2}-m_{n-1}^{2}]^{a_{n-1}} \nonumber \\
&=&\pi^{D/2}(-1)^{
{{\sum_{i=1}^{n-1}}}a_{i}}\Gamma(1+a_0)\Gamma(1+a_1)...\Gamma(1+a_{n-1}) \nonumber \\
&&\times \sum_{j_1,...,j_S}^{\infty}\frac{\Gamma(1-D/2-j_{1}-j_{2}-...-j_{w+n-1})}{j_{w+2n}!j_{w+2n+1}!...j_{S}!}\nonumber \\
&&\times \frac{(-l_1^2)^{j_{1}}}{j_{1}!}\frac{(-l_2^2)^{j_2}}{j_{2}!}...\frac{(-l_{n-1}^2)^{j_{n-1}}}{j_{n-1}!}\nonumber \\
&&\times\frac{(-l_{12}^2)^{j_n}}{j_{n}!}\frac{(-l_{13}^2)^{j_{n+1}}}{j_{n+1}!}...\frac{(-l_{n-1,n-2}^2)^{j_{w+n-1}}}{j_{w+n-1}!}\nonumber \\
&&\times\frac{(m_{0}^2)^{j_{w+n}}}{j_{w+n}!}\frac{(m_{1}^2)^{j_{w+n+1}}}{j_{w+n+1}!}...\frac{(m_{n-1}^2)^{j_{w+2n-1}}}{j_{w+2n-1}!}.
\end{eqnarray}
This expression only represent the one-loop n-point function after
the analytic continuation in the parameters $a_0,a_1,...,a_{n-1}$
to negative value.

\section{One-point function}
One-point functions at one-loop level are the simplest Feynman
loop integrals and we start with them for completeness. The
integral associated to one-loop one-point function, case $n=1$,
given by
\begin{equation}
J^{(1)}(D,\alpha_1,m)=\int d^{D}l_{0}(l_{0}^2-m^2)^{\alpha_1},
\end{equation}
can be evaluated by method above, obtain, after analytic
continuation to $i<0$, the known result

\begin{equation}
J^{(1)}(D, \alpha_1 ,m)=\pi^{D/2}(-\ai)_{-D/2}(-m^2)^{\ai+D/2}
\end{equation}
that is to according to \cite{hoof/veltman1} when $\ai=-1$, and
the Pochhammer symbol is defined as, \be (a|b) = (a)_b
=\frac{\G(a+b)}{\G(a)}, \label{poch}\ee we will turn to the left
form when we deal with pentagon and hexagon integrals, because the
number of sum indices will be large (more than 10 sometimes) and
then the first notation becomes better to read.

\section{Two-point function}
Two-point integrals are needed in order to study radiative
corrections such as self-energy and vacuum polarization. These
integrals raises no difficulties and their results are well-known
of quantum field theory courses. The integral associated to
one-loop two-point function, case $n=2$ given by
\begin{equation}
J^{(2)}(D,\ai,\aii,l_1,m_{0},m_{1})=\int
d^{D}l_{0}(l_{0}^2-m_{0}^{2})^{\ai}\left[(l_{0}-l_{1})^2-m_{1}^{2}\right]^{\aii},
\end{equation}
can be evaluated by method above, one obtains ten different
solutions analytically continued that can be calculated by general
expression
\begin{eqnarray}
J^{(2)}&=&J^{(2)}(D,\ai,\aii,l_1,m_{0},m_{1})\nonumber \\
&=&\pi^{D/2}(-1)^{\ai+\aii}\Gamma(1+\ai)\Gamma(1+\aii) \nonumber \\
&&\times \sum_{j_1,...,j_5=0}^{\infty}\frac{\Gamma(1-D/2-
j_1)}{(1)_{j_4}(1)_{j_5}}\frac{(-l_{1}^2)}{(1)_{j_1}}\frac{
(m_{0}^2)}{(1)_{j_2}}\frac{(m_{1}^2)}{(1)_{j_3}},
\end{eqnarray}
using the constraints equations
\begin{eqnarray}
D/2 &=&-j_1-j_4-j_5, \\
\ai &=&j_1+j_2+j_3,  \\
\aii &=&j_1+j_3+j_5.
\end{eqnarray}

The two point function will be obtained after the analytic
continuation of the each solution to $\ai,\aii<0$.

We choose one convenient solution given by (consider
$\sigma_2=\ai+\aii+D/2 $)

\begin{eqnarray}
J^{(2)}&=&J^{(2)}(D,\ai,\aii,l_1,m_{0},m_{1})\nonumber \\
&=&\pi^{D/2}(-m^2_1)^{\sigma_2}\{(-\aii)_{-\ai-D/2}(D/2)_{\ai}\nonumber \\
&&\times F_4[-\sigma_2,-\ai;D/2,1-\ai-D/2|\frac{l_1^2}{m_1^2};\frac{m_0^2}{m_1^2}] \nonumber \\
&&+(\frac{m_0^2}{m_1^2})^{\ai+D/2}(-\ai)_{-D/2} \nonumber \\
&&\times
F_4[-\aii,D/2;D/2,1+\ai+D/2|\frac{l_1^2}{m_1^2};\frac{m_0^2}{m_1^2}]\},
\label{b1}
\end{eqnarray}
where $F_4$ is an Appel's hypergeometric function \cite{hiperg}.
This solution contains special cases. For example when $m_0=0$,
$m_1=m$ this expression can represent the integral associated to
self energy of fermions. In this case the solution above can be
written of the form
\begin{eqnarray}
J^{(2)}&=&J^{(2)}(D,\ai,\aii,l_1,0,m) \nonumber \\
&=&\pi^{D/2}(-\aii)_{-\ai-D/2}(D/2)_{\ai}(-m^2)^{\sigma_2} \nonumber \\
&&\times _{2}F_1[-\sigma_2,-\ai;D/2|\frac{l_1^2}{m^2}],
\end{eqnarray}
where $_{2}F_1$ is the Gauss hypergeometric function
\cite{hiperg}. The analytic continuation of this expression can be
obtained by choosing of others three indices between
$j_1,...,j_5$. This solution is divergent only to great value of
$l_1 $ and the wait infrared divergence not appear here. If we
perform dimensional regularization with $D=4-2\epsilon$,
$\ai=\aii=-1 $ and $\epsilon\rightarrow 0$ in the equation above,
we get
\begin{eqnarray}
J^{(2)}&=&J^{(2)}(4-2\epsilon,-1,-1,l_1,0,m) \nonumber \\
&=&\Delta-\pi^2[\log(-m^2) \nonumber \\
&&+\log(1-\frac{l_1^2}{m^2})+\frac{m^2}{l_1^2}\log(1-\frac{l_1^2}{m^2})-1].
\end{eqnarray}
where $\Delta=\pi^{2}(\frac{1}{\epsilon}+\gamma-\log\pi)$. This
result is to according to \cite{hoof/veltman1}. Other case of
interest we have when $m_0=m_1=m$. This case is associated with
the diagram of the vacuum polarization. The expression (\ref{b1})
is then given by
\begin{eqnarray}
J^{(2)}&=&J^{(2)}(D,\ai,\aii,l_1,m,m) \nonumber \\
&=&\pi^{D/2}(-m^2)^{\sigma_2}(-\ai-\aii)_{-D/2} \times_3F_2\left[
\begin{array}{l|}
-\sigma_2,-\ai,-\aii \\
-\frac{\ai+\aii}{2},\frac{1-\ai-\aii}{2}
\end{array}
\frac{l_1^2}{4m^2}\right],\nonumber \\
&& \label{a5} \end{eqnarray} where $_{3}F_2$ is a hypergeometric
function. The above solutions as well as its analytic continuation
is to according to \cite{davyd} and \cite{suzuki-caxambu}.

\section{Three-point function}
We turn to scalar three-point functions, increasing a little bit
the difficulties, we one put the external momenta off-shell and
consider three propagators with distinct masses. The one-loop
three-point function is given by
\begin{eqnarray}
J^{(3)}&=&J^{(3)}(D,\ai,\aii,\aiii,l_1,l_2,m_{0},m_{1},m_2) \nonumber \\
&=&\int
d^{D}l_{0}[l_{0}^2-m_{0}^{2}]^{\ai}[(l_{0}-l_{1})^2-m_{1}^{2}]^{\aii}[(l_o-l_2)^2-m^2_2]^{\aiii},
\end{eqnarray}
performing the method presented in Section 1, we obtain the sixty
nine non-trivial independent solutions, all analytically
continued. We select the convenient solution (consider
$\sigma_3=\ai+\aii+\aiii+D/2$ )
\begin{eqnarray}
J^{(3)}&=&J^{(3)}(D,\ai,\aii,\aiii,l_1,l_2,m_{0},m_{1},m_2) \nonumber \\
&=&\pi^{D/2}(-m_2^2)^
{\sigma_3}(D/2)_{\ai+\aii}(-\aiii)_{-\sigma_2} \nonumber \times
\Psi_{3}\left[
\begin{array}{l|}
-\sigma_3,-\ai,-\aii \\
1-\sigma_3+\aiii,D/2
\end{array}
-\frac{l_1^2}{m_2^2};\frac{l_2^2}{m_2^2};\frac{l_{12}^2}{m_2^2};
\frac{m_0^2}{m_2^2};\frac{m_1^2}{m_2^2}\right],\nonumber \\
&& \label{a5} \end{eqnarray} where $\Psi_{3}$ is a hypergeometric
function, that can be given in terms of the generalized Lauricella
functions \cite{hiperg}, (see appendix A). With this solution we
can obtain some cases of physical interest. The case $m_0\neq 0$
and $ m_1=m_2=m$, that is applied for example in Higgs
$\rightarrow\gamma\gamma$ decay, is given by
\begin{eqnarray}
J^{(3)}&=&J^{(3)}(D,\ai,\aii,\aiii,l_1,l_2,m_{0},m) \nonumber \\
&=&\pi^{D/2}(-m^2)^{\sigma_3}\frac{(D/2)_{\ai}}{(-\sigma_3)_{\ai+D/2}} \nonumber \\
&&\times R_{11}\left[
\begin{array}{l|}
-\sigma_3,-\ai,-\aiii,-\aii \\
D/2,-\aii-\aiii,1-\ai-D/2
\end{array}
\frac{l_1^2}{m^2};\frac{m_0^2}{m^2};-\frac{l_{12}^2}{m^2};\frac{l_2^2}{m^2}\right],
\label{a6} \end{eqnarray} where $R_{11}$ is hypergeometric-type
function given in the Appendix A. If we make $m_0=0$ and $m_1\neq
m_2\neq 0$, in (\ref{a5}), we have the integral that used also in
the H decay. The expression to this solution is
\begin{eqnarray}
J^{(3)}&=&J^{(3)}(D,\ai,\aii,\aiii,l_1,l_2,m_1,m_2) \nonumber \\
&=&\pi^{D/2}(-m_2^2)^{\sigma_3}(D/2)_{\ai\aii}(-\aiii)_{-\sigma_2} \nonumber \\
&&\times R_{6}\left[
\begin{array}{l|}
-\sigma_3,-\aii,-\ai \\
D/2,1-\sigma_3+\aiii
\end{array}
\frac{l_1^2}{m_2^2};\frac{m_1^2}{m_2^2};\frac{l_{12}^2}{m_2^2};\frac{l_2^2}{m_2^2}\right],
\end{eqnarray}
where also $R_6$ is a hypergeometric function, that can be given
by generalized Lauricella functions, expressed in the Appendix A.
The others cases to one-loop three-point function, namely: 1)
$m_0=0$ and $m_1=m_2$; 2) $m_0=m_1=m_2\neq 0$; 3)$m_0=m$ and
$m_1=m_2=m$; 4) $m_0=m$ and $m_1=m_2=0$; 5) $m_0=m_1=m_2=0$, are
studied in \cite{davyd,esdras}.

\section{Four-point function}
Usually four-point integrals are the most complicated in quantum
field theory courses. They represent the scattering $(2\rightarrow
2)$ and for this reason are very important in
phenomenology\cite{duplancic}. In a previous work\cite{box} two of
us studied such integrals --- the ones that contribute to
photon-photon scattering in QED
--- and did show several hypergeometric series representing it.
Two of them were calculated before, using Mellin-Barnes approach,
by Davydychev\cite{davyd-foton}, the functions given by Appel's
$F_3$ and $\Sigma F_2$, the first one a single \hf{} and the
second a sum of four ones.

The integral associated to one-loop four-point function, case
$n=4$, is given
\begin{eqnarray}
J^{(4)}&=&J^{(4)}(D,\ai,\aii,\aiii,\aiv,l_1,l_2,l_3,m_{0},m_{1},m_2,m_3) \nonumber \\
&=&\int
d^{D}l_{0}[l_{0}^2-m_{0}^{2}]^{\ai}[(l_{0}-l_{1})^2-m_{1}^{2}]^{\aii}
[(l_o-l_2)^2-m^2_2]^{\aiii}[(l_o-l_3)^2-m^3_2]^{\aiv}, \nonumber \\
\end{eqnarray}
and is far more general than that we considered in \cite{box} and
Duplan\u ci\'c and Ni\u zi\'c presented in \cite{duplancic}.  This
integral can be evaluated by method described in the Section $1$
and we obtain one thousand and twelve non trivial solution all
analytically continued, as can be read from table-\ref{table-box}.
We choose again only the convenient solution, it is written as
(consider $\sigma_4=\ai+\aii+\aiii+\aiv+D/2$)
\begin{eqnarray}
J^{(4)}&=&J^{(4)}(D,\ai,\aii,\aiii,\aiv,l_1,l_2,l_3,m_0,m_1,m_2,m_3) \nonumber \\
&=& \pi^{D/2}(-\aiv)_{-\sigma_4+\aiv}(D/2)_{\sigma_4-\aiv-D/2}(-m_3)^{\sigma_4}\nonumber \\
&&\times \Phi\left[
\begin{array}{l|}
-\sigma_4,-\ai,-\aii,-\aiii \\
D/2,1-\sigma_4+\aiv \end{array}
\frac{-l_{1}^2}{m^2_3};\frac{-l_2^2}{m^2_3};\frac{l_3^2}{m^2_3};
\frac{-l_{12}^2}{m^2_3};\frac{l_{13}^2}{m^2_3};\frac{l_{23}^2}{m^2_3};
\frac{m_{0}^2}{m^2_3};\frac{m_{1}^2}{m^2_3};\frac{m_{2}^2}{m^2_3}\right], \nonumber \\
\end{eqnarray}
where $\Phi$ is a hypergeometric function (see Appendix A), that
can be written in terms of generalized Lauricella functions. We
observe that above hypergeometric series does not allow one to
take limit of vanishing $m_3$ since this it is not defined on that
kinematic region. In the above solution is possible take out the
limits: $m_0=m_1=m_2=0$; $m_0=m_1=0$ and $m_2\neq 0$; $m_0=0$ and
$m_1=m_2$ or $m_1\neq m_2\neq 0$; $m_0=m_1=m_2=m_3=m$; and
on-shell cases, see also \cite{box,davyd-foton}. To the special
case where $m_0=m_1=m_2=m_3=0$ we obtain the solution
\begin{eqnarray}
J^{(4)}&=&J^{(4)}(D,\ai,\aii,\aiii,\aiv,l_1,l_2,l_3) \nonumber \\
&=&\pi^{D/2}(p_3^2)^{\sigma_4}\frac{(-\ai)_{-\sigma_4}(-\aii)_{-\sigma_4}}{(\sigma_4)_{D/2}}\nonumber \\
&&\times \Psi_{3}\left[
\begin{array}{l|}
\sigma_4,-\aii,-\aiii \\
1-\sigma_4+\ai,1-\sigma_4+\aiv
\end{array}
\frac{-l_{12}^2}{l^2_3}; \frac{l_1^2}{l^2_3};\frac{l_2^2}{l^2_3};
\frac{l_{13}^2}{l^2_3};\frac{l_{23}^2}{l^2_3}\right]. \nonumber \\
\end{eqnarray}

\section{Generating functional for pentagon integral}

Pentagon integrals were studied by Melrose\cite{melrose} and
recently by Bern\cite{bern-penta} and Weinzierl and
Kosower\cite{weinzierl}, and several authors \cite{pentagonos}, in
order to study scattering process where 2 particles go to 3
particles.

Let the external legs of the pentagon be ($p_1,\; p_2-p_1,\;
p_3-p_2,\; p_4-p_3,\; p_4\;$), see figure, and let us consider
firstly the on-shell case. The generating functional for the
scalar negative-dimensional integrals is written as a special case
of the general functional of section 1, \be \label{gf-5on}G_5^{ON}
= \int d^D\! q\; \exp{\left[ -\alpha q^2-\beta(q+p_1)^2-
\gamma(q+p_2)^2-\theta(q+p_3)^2 -\omega(q+p_4)^2 \right]}, \ee
then completing the square one can easily integrate and simplify
several terms. Eventually one get, \be \label{gerf-5on} G_5^{ON} =
\left(\frac{\pi}{\lambda}\right)^{D/2} \exp{\left[
-\frac{1}{\lambda}\left( \alpha\gamma p_2^2 +\alpha\theta p_3^2
+\beta\theta s_{13} + \beta\phi s_{14} + \gamma\phi s_{24} \right)
\right] }, \ee where we define
$\lambda=\alpha+\beta+\gamma+\theta+\omega$, and three of the
Mandelstam variables for the on-shell pentagon,
\be\label{mandel-def} s_{13} = (p_1-p_3)^2,\;\;\; s_{14} =
(p_1-p_4)^2,\;\;\; s_{24} = (p_2-p_4)^2. \ee

Following the usual procedure applied in NDIM we count the total
number of solutions we will have to deal with. Exponential gives
us five sums (Taylor expansion for each argument), $\lambda$ other
five (multinomial expansion), and the equations are the number of
propagators plus one, six. Then, there are $C_{10,6}=210$ possible
ways to solve the $6\times 6$ system. The ones that have null
determinant do not have solution, they are 85, the remaining 125
after properly solved will provide us hypergeometric series
representing the original Feynman integral.

On the other hand, projecting out powers of the exponential in
(\ref{gf-5on}), \be G_5^{ON} =
\sum_{\alpha_1,...\alpha_5=0}^\infty
\frac{(-1)^{\ai+\aii+\aiii+\aiv+\av} \alpha^{\ai}\beta^{\aii}
\gamma^{\aiii} \theta^{\aiv}
\omega^{\av}}{\ai!\aii!\aiii!\aiv!\av!}{\cal
J}_5^{ON}(\ai,\aii,\aiii,\aiv,\av), \ee where \be {\cal
J}_5^{ON}(\ai,\aii,\aiii,\aiv,\av) = \int d^D\! q\; (q^2)^{\ai}
(q+p_1)^{2\aii} (q+p_2)^{2\aiii} (q+p_3)^{2\aiv} (q+p_4)^{2\av},
\ee is the negative-dimensional integral.

We will present two kinds of \hf{}s since several others can be
written immediately if one observe the symmetries of the diagram
(or alternatively the symmetries of the generating functional),

\beq && \left\{ \aiii\leftrightarrow \aiv, p_1\lra p_4-p_3,
p_2-p_1\lra p_4 \right\}, \left\{ \av\leftrightarrow \aiv, p_1\lra
p_3-p_2, p_4\lra p_2-p_1 \right\},\nonumber\\ && \left\{
\ai\leftrightarrow \av, p_1\lra p_3-p_2, p_2-p_1\lra p_4-p_3
\right\}, \eeq and so forth, there are several other symmetries of
the integral (\ref{gerf-5on}). When we have a hypergeometric
series one can obtain some others merely interchanging exponents
of propagators and external momenta, where three such sets are
given in the above list.

\subsection{Hypergeometric functions for massless on-shell pentagon}

Let us define $\s_5=\alpha_{12345}+D/2$ and the shorthand notation
we will use hereafter in order to have a more compact notation,

$$j_{ab} = j_a+j_b$$ and so on. Observe that we use this compact
notation only for sum index. Mandelstam variables are given by
(\ref{mandel-def}).

The first \hf{} representing the Feynman loop integral is given
\be {\cal J}_5^{ON}(\{\alpha_i\}) = \G_5^{(1)} {\cal
S}^{(5)}_4\left[
\begin{array}{l|} -\ai, -\aii, -\av,-\s_5-D/2 \\
1+\aiii-\s_5, 1+\av-\s_5\end{array} \frac{p_{2}^2}{s_{24}},
-\frac{p_{3}^2}{s_{24}}, \frac{s_{14}}{s_{24}},
-\frac{s_{13}}{s_{24}}\right], \label{penta-on1}\ee the possible
poles are contained in the factor \be \G_5^{(1)} = \pi^{D/2}
(-\aiii|\s_5)(-\av|\s_5) (\s_5+D/2|-2\s_5-D/2) s_{24}^{\s_5}, \ee
where the Pochhammer symbol is given by eq.(\ref{poch}). In the
appendix we define all hypergeometric functions we use in this paper.

The above \hf{} is the "symmetric" solution calculated by
Davydychev\cite{davyd}, using Mellin-Barnes integral
representation approach.

However, in the negative-dimensional approach we obtain, in
general, several hypergeometric functions, which represent it and
are related by \ac{} (direct or indirect). We write down other
\hf{} for the massless scalar pentagon, \beq {\cal
J}_5^{ON}(\{\alpha_i\}) &=& \G_5^{(2)} {\cal T}^{(5)}_4\left[
\begin{array}{l|} -\ai, -\aiii, -\aiv, D/2+\alpha_{123} \\
1+\alpha_{123}+D/2, 1-\alpha_{145}-D/2\end{array}
-\frac{p_{2}^2}{s_{24}}, \frac{p_{3}^2}{s_{13}},
-\frac{s_{14}}{s_{24}}, \frac{s_{14}}{s_{13}}\right],
\label{penta-on2} \eeq

where \be \G_5^{(2)} = \pi^{D/2} (-\aii|\s_5-k)(-\av|\s_5-m)
(\s_5+D/2|-2\s_5-D/2+\alpha_{34})
\frac{s_{14}^{\s_5-\alpha_{34}}}{s_{13}^{-\aiv} s_{24}^{-\aiii}},
\ee the same set of permutations of exponents of propagators and
external momenta can be applied to the above result, in order to
generate several others. 


\subsection{Hypergeometric functions for massive on-shell pentagon}
We now turn to the case where the five propagators have masses,
and let them to be distinct. We will present a hypergeometric
function that allows us to consider interesting particular cases:
2, 3, 4 and 5 equal masses.

When we deal with massive propagators the system we must solve is
greater than the related to the massless diagram. Let the masses
be $m_1^2$ attached to the propagators labelled as $\ai$, $m_2^2$
attached to the one labelled as $\aii$ and so on.

The generating functional is simply the massless one
(\ref{gerf-5on}) times the exponential containing the masses, \be
\label{gf-5on-5m} G^{ON}_{5M} = G^{ON}_{5} \exp{\left(\alpha m_1^2
+\beta m_2^2+\gamma m_3^2 +\theta m_4^2 +\omega m_5^2\right) }
,\ee the total number of systems we must solve now is given by the
combinatorics $C_{15,6}=5005$. Among them we pick the most
convenient one, i.e., the hypergeometric series that will allow us
to study several important special cases. We choose to present the
result

\beq {\cal J}_5^{ON}(m_i^2) &=& \G_5^{({\rm mass})}(\av;m_5) {\cal
S}^{(5)}_9\left[
\begin{array}{l|} -\ai, -\aii,-\aiii,-\aiv,-\s_5 \\
D/2, 1+\av-\s_5\end{array} z_1;...;z_{9}\right],\label{penta-5m}
\eeq where \beq && z_1=\frac{p_{2}^2}{m_{5}^2},\;\;
z_2=\frac{p_{3}^2}{m_{5}^2},\;\; z_3=-\frac{s_{14}}{m_{5}^2},\;\;
z_4=\frac{s_{13}}{m_{5}^2},\;\; z_5=-\frac{s_{24}}{m_{5}^2},\;\;
z_6=\frac{m_{1}^2}{m_{5}^2},\;\; z_7= \frac{m_{2}^2}{m_{5}^2},\;\;
\nonumber\\
&& z_8=\frac{m_{3}^2}{m_{5}^2},\;\; z_9=\frac{m_{4}^2}{m_{5}^2},
\eeq which is valid for five different masses, such that $m_5\neq
0$ and we define \be\label{gama5mass} \G_5^{({\rm mass})}(\av;m_5)
= \pi^{D/2} (-\av|\s_5) (D/2|-\s_5-D/2) (m_5^2)^{\s_5}, \ee
observe that the above 9-fold hypergeometric series allows us to
study the cases where any of the masses $m_1,m_2,m_3,m_4$ vanish,
or any of the cases where they are equal to $m_5$ (which in this
kinematical region can not vanish).

We proceed to special cases now. First, let $m_1=m_5$. Then, the
hypergeometric series in the $l_1$ index can be rewritten as
$_2F_1(...|1)$. Using Gauss' summation formula\cite{hiperg}, \be
_2F_1(a,b,c|1) = \frac{\G(c)\G(c-a-b)}{\G(c-a)\G(c-b)},\ee we can
exactly sum up the series. Rewriting the Pochhammer symbols
involving $l_1$ as, \beq\label{soma-z1} \sum_{l_1=0}^\infty
\frac{(-\ai|j_{12}+l_1)
(-\s_5|j_{12345}+l_{1234})}{l_1!(1+\av-\s_5|j_{124}+l_{1234})} &=&
\frac{ (-\ai|j_{12}) (-\s_5|j_{12345}+l_{234})
}{(1+\av-\s_5|j_{124}+l_{234})
}\\
&&\times \sum_{l_1=0}^\infty  \frac{(-\s_5+j_{12345}+l_{234}|l_1)
(-\ai+j_{12}|l_1)}{l_1!(1+\av-\s_5+j_{124}+l_{234}|l_1)},\nonumber\eeq
where we used the property of Pochhammer symbols, \be (a|b+c) =
(a|b)(a+b|c), \ee thus the series in $l_1$ is recast as a $_2F_1$
of unity argument, and then can be summed, \be \sum_{l_1=0}^\infty
\left[\frac{(|)...}{(|)...} \right] =
\frac{(-\ai|j_{12})(-\s_5|j_{12345}+l_{234})
(A_1|-j_{1235})}{(A_2|-j_{35})(A_3|j_4+l_{234})}
\frac{\G(1+\av-\s_5)\G(A_1)}{\G(A_2)\G(A_3)} ,\ee where \be
A_1=1+\alpha_{15},\qquad A_2=1+\av,\qquad A_3=1+\alpha_{15}-\s_5,
\ee the same procedure of summing up series of hypergeometric type
was applied in \cite{lab-test}.

Substituting the above result in (\ref{penta-5m}) one obtain the
result for the pentagon integral having four different masses is
given by,

\beq {\cal J}_5^{ON}(m_i^2;4) &=& \G_5^{({\rm
mass})}(\alpha_{15};m_5) {\cal S}^{(5)}_8\left[
\begin{array}{l|}-\ai, -\aii,-\aiii,-\aiv,-\av, -\s_5 \\
D/2, -\alpha_{15},1+\alpha_{15}-\s_5\end{array}
z_1;...;z_{8}\right],\eeq where \beq &&
z_1=-\frac{p_{2}^2}{m_{5}^2},\;\;
z_2=-\frac{p_{3}^2}{m_{5}^2},\;\; z_3=-\frac{s_{14}}{m_{5}^2},\;\;
z_4=\frac{s_{13}}{m_{5}^2},\;\; z_5=-\frac{s_{24}}{m_{5}^2},\;\;
z_6=\frac{m_{2}^2}{m_{5}^2},\;\;
z_7=\frac{m_{3}^2}{m_{5}^2},\nonumber\\ &&
z_8=\frac{m_{4}^2}{m_{5}^2} ,\eeq where the argument "4" is to
remember one the number of different masses and the factor
involving gamma functions, given by (\ref{gama5mass}), is greatly
simplified, and we perform the \ac{} in the final step of the
calculation (sum the series and then analytically continue the
Pochhammer symbols).

The same algebraic manipulations can be made to obtain the cases
where the pentagon has three different masses $m_5,m_4,m_3$, \beq
{\cal J}_5^{ON}(m_i^2;3) &=& \G_5^{({\rm mass})}(\alpha_{125};m_5)
{\cal S}^{(5)}_7\left[
\begin{array}{l|} -\ai,-\aii,-\aiii,-\aiv,-\av,-\s_5 \\
D/2, -\alpha_{125}, 1+\alpha_{125}-\s_5\end{array}
z_1;...;z_{7}\right],\eeq where the variables are, \be
z_1=-\frac{p_{2}^2}{m_{5}^2},\;\;
z_2=-\frac{p_{3}^2}{m_{5}^2},\;\; z_3=\frac{s_{14}}{m_{5}^2},\;\;
z_4=-\frac{s_{13}}{m_{5}^2},\;\; z_5=-\frac{s_{24}}{m_{5}^2},\;\;
z_6=\frac{m_{3}^2}{m_{5}^2},\;\; z_7=\frac{m_{4}^2}{m_{5}^2}.\ee

When only two masses are different we can sum the $l_3$ index,
\beq {\cal J}_5^{ON}(m_i^2;2) &=& \G_5^{({\rm
mass})}(\alpha_{1235};m_5) {\cal S}^{(5)}_6\left[
\begin{array}{l|} -\ai,-\aii,-\aiii,-\aiv,-\av,-\s_5 \\
D/2, -\alpha_{1235}, 1+\alpha{1235}-\s_5\end{array}
z_1;...;z_{6}\right], \eeq where its six variables are, \be
z_1=\frac{p_{2}^2}{m_{5}^2},\;\; z_2=-\frac{p_{3}^2}{m_{5}^2},\;\;
z_3=\frac{s_{14}}{m_{5}^2},\;\; z_4=-\frac{s_{13}}{m_{5}^2},\;\;
z_5=\frac{s_{24}}{m_{5}^2},\;\; z_6=\frac{m_{4}^2}{m_{5}^2}, \ee
finally, the very special case where all five masses are equal,
\beq {\cal J}_5^{ON}(m_i^2;1) &=& \pi^{D/2}(m_5^2)^{\s_5}
(-\s_5+D/2|-D/2) \nonumber\\
&& \times {\cal S}^{(5)}_5\left[
\begin{array}{l|}-\ai,-\aii,-\aiii,-\aiv,-\av,-\s_5 \\
-\alpha_{12345}\end{array} z_1;...;z_{5}\right], \eeq where, \be
z_1=-\frac{p_{2}^2}{m_{5}^2},\;\; z_2=\frac{p_{3}^2}{m_{5}^2},\;\;
z_3=-\frac{s_{14}}{m_{5}^2},\;\; z_4=\frac{s_{13}}{m_{5}^2},\;\;
z_5=-\frac{s_{24}}{m_{5}^2},\ee observe that two Pochhammer
symbols cancelled. The above result is a generalization of our
previous study of box integrals pertaining to photon-photon
scatering\cite{box}.

\subsection{Hypergeometric functions for massless off-shell pentagon}

Massless pentagon integrals were necessary to Bern, Dixon and
Kosower\cite{bern-penta} in order to study $2\rightarrow 3$
scattering, such as $e^+e^- \rightarrow {\rm 3\; jets}$. In that
work two of the external momenta were considered to be massive and
the remaining ones massless.

Here we will present the whole five possibilities: from one to
five massive external particles. In our approach the difficulty to
carry out the integrals with different masses or equal ones is the
same. So, we prefer the most general case, five distinct masses.

Hereafter we consider massive external particles, \be
p_1^2=M^2,\quad (p_1-p_2)^2 = M_{12}^2, \quad
(p_2-p_3)^2=M_{23}^2, \quad (p_3-p_4)^2 = M_{34}^2, \quad
p_4^2=M_4^2. \ee

\subsubsection{One external leg off-shell}

We present two four-fold hypergeometric series representing the
scalar pentagon integral with one massive external particle.

\beq {\cal J}_5^{(1)}(\{\alpha_i\}) &=& \G_5(\ai,\aii;M^2) {\cal
S}^{(5-1off)}_5\left[
\begin{array}{l|}-\aiii,-\aiv,-\av,-\s_5 \\
1+\ai-\s_5, 1+\aii-\s_5\end{array} z_1;...;z_{5}\right],\eeq where
its five variables are, \be z_1=\frac{p_2^2}{M^2},\;\; z_2=
\frac{p_3^2}{M^2},\;\; z_3= \frac{s_{13}}{M^2},\;\; z_4=
\frac{s_{14}}{M^2},\;\; z_5= -\frac{s_{24}}{M^2}, \ee where the
superscript $(1)$ means one leg off-shell and \be\label{gama5-ij}
\G_5(\ai,\aii;M^2) = \pi^{D/2} (-\ai|\s_5)(-\aii|\s_5)
(\s_5+D/2|-2\s_5-D/2) (M^2)^{\s_5}, \ee observe that the above
hypergeometric series {\it does not allow} one to take the limit
of vanishing $M$ since this it is not defined on that point.

However, there are others multiple series that allow us to take
such limit. One example is, \beq {\cal J}_5^{(1)}(\{\alpha_i\})
&=& \G_5(\aii,\aiv;s_{13}) {\cal T}^{(5-1off)}_5\left[
\begin{array}{l|}-\ai,-\aiii,-\av,-\s_5 \\
1+\aii-\s_5, 1+\aiv-\s_5\end{array}
z_1;...;z_{5}\right],\label{1-off-penta}\eeq define, \be
z_1=-\frac{p_2^2}{s_{13}},\;\; z_2=\frac{p_3^2}{s_{13}},\;\; z_3=
\frac{s_{14}}{s_{13}},\;\; z_4= -\frac{s_{24}}{s_{13}},\;\; z_5=
\frac{M^2}{s_{13}},\ee in the limit of $M=0$ the series on $j_6$
index reduces to its first term, unity, so we are left with a
four-fold hypergeometric series.

\subsubsection{Two external legs off-shell}
Next we put another external leg off mass shell, namely, one
consider $(p_1-p_2)^2=M_{12}^2$. The first hypergeometric series
we collect from the total 924 ones (see table \ref{table-penta}) ,
is

\beq {\cal J}_5^{(2)}(\ai,\aii,\aiii,\aiv,\av) &=&
\G_5(\ai,\aii;M^2) {\cal S}^{(5-2off)}_6\left[
\begin{array}{l|}-\aiii,-\aiv,-\av,-\s_5 \\
1+\ai-\s_5, 1+\aii-\s_5\end{array} z_1;...;z_{6}\right], \eeq its
six variables are given by, \be z_1=\frac{p_2^2}{M^2},\;\; z_2=
\frac{p_3^2}{M^2},\;\; z_3= \frac{s_{13}}{M^2},\;\; z_4=
-\frac{s_{14}}{M^2},\;\; z_5= \frac{s_{24}}{M^2},\;\; z_6=
\frac{M_{12}^2}{M^2},\ee where the superscript $(2)$ means two
legs off-shell. Observe again that the above hypergeometric series
does not allow one to take the limit of vanishing $M$ but admits
the limit of $M_{12}=0$ to be taken.

We can consider also the case where the two masses are equal. Then
the hypergeometric function in the $j_7$ index reduces to a
gaussian one, namely $_2F_1$, that can be exactly summed when its
argument is unity.

Another 6-fold hypergeometric series \beq {\cal
J}_5^{(2)}(\ai,\aii,\aiii,\aiv,\av) &=& \G_5'
(\ai,\aii,\aiii,;M_{12}^2/p_2^2) \nonumber\\
&&\times {\cal T}^{(5-2off)}_6\left[
\begin{array}{l|}\aii-\s_5,-\aii,-\aiv,-\av, D/2+\alpha_{345} \\
1+\alpha_3-\s_5\end{array} z_1;...;z_{6}\right], \eeq where \be
z_1= \frac{p_3^2}{p_2^2},\;\; z_2= \frac{s_{13}}{M_{12}^2}
z_3=\frac{s_{14}}{M_{12}^2},\;\; z_4= -\frac{s_{24}}{p^2_2},\;\;
z_5= \frac{M^2}{M_{12}^2},\;\; z_6= -\frac{p^2_2}{M_{12}^2}, \ee
where the factor is defined by, \be
\G_5'(\ai,\aii,\aiii,;M_{12}^2/p_2^2) = (-\ai|\s_5-\aii)
(-\aiii|\s_5) (\s+D/2|-2\s_5-D/2+\aii) \left(
\frac{M_{12}^2}{p_2^2}\right)^{\aii} (p_2^2)^{\s_5} , \ee taking
$M_{12}=M$ we can sum up, using Gauss' summation
formula\cite{hiperg}, the $j_6$ series and get,

\beq {\cal J}_5^{(2)}(\ai,\aii,\aiii,\aiv,\av) &=&
\G_M\G_5'(\ai,\aii,\aiii,;M^2/p_2^2) \nonumber\\ &&\times {\cal
U}^{(5-2off)}_5\left[
\begin{array}{l|}\aii-\s_5,-\aii,-\aiv,-\av, D/2+\alpha_{345} \\
1+\alpha_{23}-\s_5, 1-\alpha_{45}-\s_5-D/2\end{array}
z_1;...;z_{5}\right], \eeq where \be z_1=\frac{p_3^2}{p_2^2},\;\;
z_2= \frac{s_{13}}{M^2},\;\; z_3= \frac{s_{14}}{M^2},\;\; z_4=
-\frac{s_{24}}{p^2_2},\;\; z_5= -\frac{p^2_2}{M^2} ,\ee and \be
\G_M = \frac{1}{(-\aiii+\s_5|-\aii)
(\s_5+\alpha_{45}+D/2|-\aii)}.\ee

\subsubsection{Three external legs off-shell}
Continue to study the off-shell pentagon, now where three external
particles are massive, the third one being $(p_2-p_3)^2=M_{23}^2$,
and one such hypergeometric series representation,

\beq {\cal J}_5^{(3)}(\ai,\aii,\aiii,\aiv,\av) &=&
\G_5(\aiii,\aiv;M^2) \nonumber\\ &&\times {\cal
S}^{(5-3off)}_7\left[
\begin{array}{l|}-\ai,-\aii,-\av,-\s_5 \\
1+\alpha_{3}-\s_5, 1-\alpha_{4}-\s_5\end{array}
z_1;...;z_{7}\right], \eeq where \be z_1=
\frac{p_2^2}{M_{23}^2},\;\; z_2= \frac{p_3^2}{M_{23}^2},\;\; z_3=
\frac{s_{13}}{M_{23}^2},\;\; z_4= -\frac{s_{14}}{M_{23}^2},\;\;
z_5=\frac{s_{24}}{M_{23}^2},\;\; z_6= -\frac{M^2}{M_{23}^2},\;\;
z_7= \frac{M_{12}^2}{M_{23}^2}, \ee that also admits one to take
two masses to vanish. Observe that $M_{23}$ can never be zero.

\subsubsection{Four external legs off-shell}
The last step before we put all five legs off mass shell is to
make $(p_3-p_4)^2=M_{34}^2$. Then we present two sample series:
the first is suitable for equal masses limit but non-vanishing,
\beq\label{4off} {\cal J}_5^{(4)}(\{\alpha_i\}) &=&
\G_5(\ai,\aii;M^2) {\cal S}^{(5-4off)}_8\left[
\begin{array}{l|}-\aiii,-\aiv,-\av,-\s_5 \\
1+\alpha_{1}-\s_5, 1-\alpha_{2}-\s_5 \end{array}
z_1;...;z_{8}\right], \eeq where its variables are, \be
z_1=\frac{p_2^2}{M^2},\;\; z_2= \frac{p_3^2}{M^2},\;\; z_3=
\frac{s_{13}}{M^2},\;\; z_4= \frac{s_{14}}{M^2},\;\; z_5=
-\frac{s_{24}}{M^2},\;\; z_6= \frac{M_{12}^2}{M^2},\;\; z_7=
-\frac{M_{23}^2}{M^2},\;\; z_8= -\frac{M_{34}^2}{M^2} ,\ee and the
second one suitable for vanishing masses limit, \beq\label{4on}
{\cal J}_5^{(4)}(\{\alpha_i\}) &=& \G_5(\aiii,\av;s_{24}) {\cal
T}^{(5-4off)}_8\left[
\begin{array}{l|}-\ai,-\aii,-\aiv,-\s_5 \\
1+\alpha_{3}-\s_5, 1-\alpha_{5}-\s_5 \end{array}
z_1;...;z_{8}\right], \eeq being \beq &&
z_1=\frac{p_2^2}{s_{24}},\;\; z_2= -\frac{p_3^2}{s_{24}},\;\; z_3=
-\frac{s_{13}}{s_{24}},\;\; z_4= \frac{s_{14}}{s_{24}},\;\; z_5=
-\frac{M^2}{s_{24}},\;\; z_6= \frac{M_{12}^2}{s_{24}},\;\; z_7=
\frac{M_{23}^2}{s_{24}}, \nonumber\\ &&
z_8=\frac{M_{34}^2}{s_{24}} ,\eeq or when the masses are such that
$s_{24} >> M^2_j$.

Note that the structure of the above series is similar but not
equal. In the first one, only one index $(j_9)$ appears five times
(in other words, the series in $j_9$ can be rewritten as a $_3F_2$
function) the other eight indices appear only three times. The
second 9-fold hypergeometric series has the indices $j_3$ and
$j_6$ appearing five times, what turns it to be more complicated
than the former. They (equations (\ref{4off}) and (\ref{4on}) )
are very similar but are not the same 9-fold series.

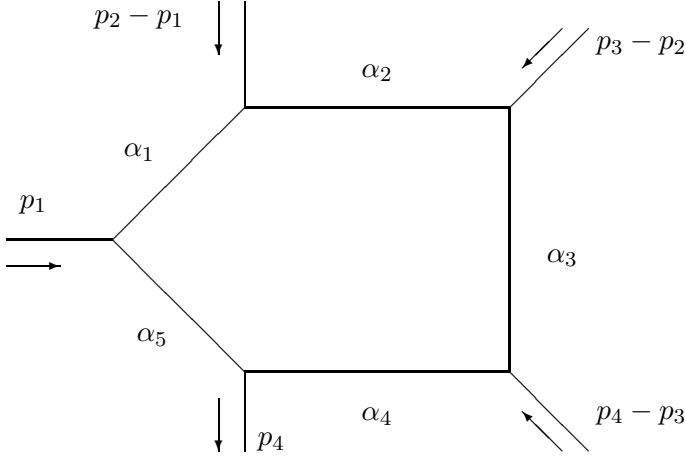
\begin{figure}
\begin{center}
\begin{picture}(600,200)(0,150)
\vspace{40mm} \thinlines

\put(180,200){\line(1,0){100}} 
\put(180,200){\line(-1,1){50}} 
\put(130,250){\line(1,1){50}} 
\put(180,300){\line(1,0){100}} 
\put(280,300){\line(0,-1){100}} 

\put(190,170){\makebox(0,0)[b]{$p_4$}}
\put(170,190){\vector(0,-1){20}} \put(180,200){\line(0,-1){30}}

{\small

\put(330,320){\makebox(0,0)[b]{$p_3-p_2$}}
\put(300,330){\vector(-1,-1){15}} \put(310,330){\line(-1,-1){30}}

\put(140,330){\makebox(0,0)[b]{$p_2-p_1$}}
\put(180,300){\line(0,1){40}} \put(170,340){\vector(0,-1){20}}

\put(130,250){\line(-1,0){40}} 
\put(100,260){\makebox(0,0)[b]{$p_1$}}
\put(90,240){\vector(1,0){20}}

\put(280,200){\line(1,-1){30}} 
\put(330,180){\makebox(0,0)[b]{$p_4-p_3$}}
\put(300,170){\vector(-1,1){15}}

\put(230,310){\makebox(0,0)[b]{$\aii$}}
\put(145,210){\makebox(0,0)[b]{$\av$}}
\put(230,180){\makebox(0,0)[b]{$\aiv$}}
\put(140,280){\makebox(0,0)[b]{$\ai$}}
\put(300,240){\makebox(0,0)[b]{$\aiii$}}}
\end{picture}\caption{Scalar massless pentagon. Labels $(i,j,k,l,m)$ represent exponents of propagators and the arrows
show momentum flow. In the case where the external legs are
off-shell we define $p_1^2=M^2$, $s_{12}=M^2_{12}$,
$s_{23}=M^2_{23}$, $s_{34}=M^2_{34}$, and $p_4^2=M^2_4$.}
\end{center}
\end{figure}

\subsubsection{Five external legs off-shell}

Now we turn to the general case where all five legs are off-shell.
Despite there are a great number of hypergeometric 9-fold series,
see table \ref{table-penta}, we present here just two of them,

\beq\label{5off1} {\cal J}_5^{(5)}(\{\alpha_i\}) &=&
\G_5(\ai,\av;M_4^2) {\cal S}^{(5-5off)}_9\left[
\begin{array}{l|}-\aii,-\aiii,-\aiv,-\s_5 \\
1+\ai-\s_5, 1+\av-\s_5 \end{array} z_1;...;z_{9}\right], \eeq its
nine variables are given as follows, \beq &&
z_1=\frac{p_2^2}{M_{4}^2},\;\; z_2= \frac{p_3^2}{M_{4}^2},\;\; z_=
-\frac{s_{13}}{M_{4}^2},\;\; z_3= \frac{s_{14}}{M_{4}^2},\;\; z_4=
\frac{s_{24}}{M_{4}^2},\;\; z_5= \frac{M^2}{M_{4}^2},\;\; z_6=
-\frac{M_{12}^2}{M_{4}^2}, \nonumber\\ &&  z_7=
-\frac{M_{23}^2}{M_{4}^2},\;\; z_8= \frac{M_{34}^2}{M_{4}^2} ,\eeq
other hypergeometric 9-fold series that have two masses in the
denominator \beq\label{5off2} {\cal
J}_5^{(5)}(\ai,\aii,\aiii,\aiv,\av) &=&
\G_5'(\aiv,\ai,\av;M_4^2/M_{34}^2) \nonumber\\ && \times {\cal
T}^{(5-5off)}_9\left[
\begin{array}{l|}-\ai,-\aii,-\aiii,\ai-\s_5 \\
1+\alpha_{14}-\s_5, 1-\av+\s_5 \end{array} z_1;...;z_{9}\right],
\eeq where we define, \beq && z_1=\frac{p_2^2}{M_{4}^2},\;\; z_2=
-\frac{p_3^2}{M_{4}^2},\;\; z_3= \frac{s_{13}}{M_{34}^2},\;\; z_4=
\frac{s_{14}}{M_{34}^2},\;\; z_5= \frac{s_{24}}{M_{34}^2},\;\;
z_6= \frac{M^2}{M_{4}^2}, \nonumber\\&& z_7=
-\frac{M_{12}^2}{M_{34}^2},\;\; z_8=
\frac{M_{23}^2}{M_{34}^2},\;\; z_9= \frac{M_{34}^2}{M_{4}^2} .\eeq

The last we present is a hypergeometric series that allows one to
take the limit of vanishing masses (any of them),

\beq\label{5off3} {\cal J}_5^{(5)}(\{\alpha_i\}) &=&
\G_5'(\av,\ai,\aiii;p_2^2/s_{24}) \nonumber\\ && \times {\cal
U}^{(5-5off)}_9\left[
\begin{array}{l|}\ai-\s_5,-\ai,-\aii,\aiv \\
1+\alpha_{3}-\s_5, 1-\alpha_{15}+\s_5  \end{array}
z_1;...;z_{9}\right],\eeq where \beq &&
z_1=\frac{p_3^2}{p_2^2},\;\; z_2= -\frac{s_{13}}{s_{24}},\;\; z_3=
\frac{s_{14}}{s_{24}},\;\; z_4= \frac{M^2}{p_{2}^2},\;\; z_5=
\frac{M_{12}^2}{s_{24}},\;\; z_6= \frac{M^2_{23}}{s_{24}},\;\;
z_7= \frac{M_{34}^2}{s_{24}},\nonumber\\ && z_8=
-\frac{M_{4}^2}{p_{2}^2},\;\; z_9= -\frac{s_{24}^2}{p_{2}^2} .\eeq

In the negative-dimensional approach we can select the kinematical
region one would like to study and then work with hypergeometric
series defined on that region. {\it All hypergeometric series
presented in this paper are exact, there are no approximations.}
They converge very fast as we have verified in \cite{nonplanar}, a
precision of 20 digits can be achieved very quickly.

\section{Hexagon Feynman loop integral}
Recently Bern, Dixon and Kosower\cite{bern-6pernas} studied
amplitudes of process involving six particles, namely,
$e^+e^-\rightarrow 4{\rm partons}$. Also, QCD corrections to
$e^+e^-\rightarrow 4{\rm jets}$ were calculated by Weinzierl and
Kosower\cite{weinzierl}; Binoth, Guillet, Heinrich and Schubert
\cite{binoth} show how to reduce the hexagon diagrams in order to
study Yukawa model at one-loop level. Their works motivates us to
perform such integrals in general cases, i.e., arbitrary exponents
of propagators, massive external legs and propagators.

The general formula we calculated in section I gives us the
generating functional,

\beq G_{Hex} &=& \left(\frac{\pi}{\zeta}\right)^{D/2} \exp{\left[
-\frac{1}{\zeta}\left( \alpha\gamma p_2^2 +\alpha\theta p_3^2
+\alpha\phi p_4^2+\beta\theta s_{13} + \beta\phi s_{14}
+\beta\omega s_{51} + \gamma\phi s_{24}  +\gamma\omega s_{25}
\right.\right.} \nonumber\\
&& \left.\left. +\theta\omega s_{35} \right) \right], \eeq define
$\s_6=\alpha_{123456}+D/2$ and $\zeta=\lambda+\phi$. Total number
of systems $C_{15,7}=6435$, such that, 2790 of them have
solutions, the remaining 3645 do not have interest at all.

The first series we write is of a kind
Davydychev\cite{davyd,davyd2} called "symmetric", \beq\label{6on1}
{\cal J}_6^{(0)}(\{\alpha_i\}) &=& \G_6(\aiv,\avi;s_{35}) {\cal
S}^{(6-on)}_8\left[
\begin{array}{l|}-\ai,-\aii,-\aiii,\av,-\s_6 \\
1+\aiv-\s_6, 1+\avi-\s_6 \end{array} z_1;...;z_{8}\right], \eeq
where the function has eight variables, \beq &&
z_1=-\frac{p_2^2}{s_{35}},\;\; z_2= \frac{p_{3}^2}{s_{35}},\;\;
z_3= -\frac{p_{4}^2}{s_{35}},\;\; z_4= \frac{s_{13}}{s_{35}},\;\;
z_5= -\frac{s_{14}}{s_{35}},\;\; z_6= \frac{s_{51}}{s_{35}},\;\;
z_7= -\frac{s_{24}}{s_{35}},\nonumber\\ && z_8=
\frac{s_{35}}{s_{35}} ,\eeq the superscript $(0)$ means that no
external leg is off mass shell. In the appendix we define all
hypergeometric functions we use in this paper in terms of
generalized Lauricella functions.

The second one pertain to another kind, another kinematical
region, \beq\label{6on2} {\cal J}_6^{(0)}(\{\alpha_i\}) &=&
\G_6(\avi,\av,\aiii;s_{24}/s_{25}) {\cal T}^{(6-on)}_8\left[
\begin{array}{l|}\av-\s_6,-\ai,-\aii,\aiv,-\av \\
1+\aiii-\s_6, 1+\alpha_{56}-\s_6 \end{array} z_1;...;z_{8}\right],
\eeq define, \beq && z_1=\frac{p_2^2}{s_{25}},\;\;
z_2=-\frac{p_{3}^2}{s_{25}},\;\; z_3= \frac{p_{4}^2}{s_{24}},\;\;
z_4= -\frac{s_{13}}{s_{25}},\;\; z_5= \frac{s_{14}}{s_{24}},\;\;
z_6= \frac{s_{51}}{s_{25}},\;\; z_7= \frac{s_{35}}{s_{25}},
\nonumber\\&& z_8= \frac{s_{25}}{s_{24}} ,\eeq where in both
hypergeometric series the factor $\G_6$ means that in equation
(\ref{gama5-ij}) we change $\s_5$ by $\s_6$.

\begin{figure}
\begin{center}
\begin{picture}(600,200)(0,150)
\vspace{40mm} \thinlines

\put(180,200){\line(1,0){100}} 
\put(180,200){\line(-1,1){50}} 
\put(130,250){\line(1,1){50}} 
\put(180,300){\line(1,0){100}} 
\put(280,300){\line(1,-1){50}} 
\put(330,250){\line(-1,-1){50}} 

\put(190,170){\makebox(0,0)[b]{$p_5$}}
\put(170,180){\vector(0,-1){20}} \put(180,200){\line(0,-1){40}}

{\small

\put(310,330){\makebox(0,0)[b]{$p_3-p_2$}}
\put(270,340){\vector(0,-1){20}} \put(280,300){\line(0,1){40}}

\put(140,330){\makebox(0,0)[b]{$p_2-p_1$}}
\put(180,300){\line(0,1){40}} \put(170,340){\vector(0,-1){20}}

\put(130,250){\line(-1,0){40}} 
\put(100,260){\makebox(0,0)[b]{$p_1$}}
\put(90,240){\vector(1,0){20}}

\put(280,200){\line(0,-1){40}} 
\put(310,170){\makebox(0,0)[b]{$p_5-p_4$}}
\put(270,160){\vector(0,1){20}}

\put(330,250){\line(1,0){40}} 
\put(350,260){\makebox(0,0)[b]{$p_4-p_3$}}
\put(370,240){\vector(-1,0){20}}

\put(230,310){\makebox(0,0)[b]{$\aii$}}
\put(145,210){\makebox(0,0)[b]{$\avi$}}
\put(230,180){\makebox(0,0)[b]{$\av$}}
\put(140,280){\makebox(0,0)[b]{$\ai$}}
\put(320,220){\makebox(0,0)[b]{$\aiv$}}
\put(310,280){\makebox(0,0)[b]{$\aiii$}}}
\end{picture}\caption{Scalar massless hexagon. Labels $(\ai,\aii,\aiii,\aiv,\av,\avi)$ represent
exponents of propagators and the arrows
show momentum flow. In the case where the external legs are
off-shell we define $p_1^2=M^2$, $s_{12}=M^2_{12}$,
$s_{23}=M^2_{23}$, $s_{34}=M^2_{34}$, $s_{45}=M^2_{45}$ and
$p_5^2=M^2_5$.}
\end{center}
\end{figure}

\subsection{One massive external leg}
Considering one of the external legs of the hexagon to be off mass
shell means that one must to deal with almost two times more
systems of algebraic equations, see table (\ref{table-hex}).

However, some hypergeometric series follow some pattern and we
think it can be generalized for scalar $N$-point functions.

The first one we write does not allow one to take the mass to
vanish, \beq\label{6off1a} {\cal J}_6^{(1)}(\{\alpha_i\}) &=&
\G_6(\ai,\aii;M_1^2) {\cal S}^{(6-1off)}_9\left[
\begin{array}{l|}-\aiii,-\aiv,-\av,\avi,-\s_6   \\
1+\ai-\s_6, 1+\aii-\s_6 \end{array} z_1;...;z_{9}\right], \eeq
where \beq && z_1=\frac{p_2^2}{M_1^2},\;\; z_2=
\frac{p_{3}^2}{M_1^2},\;\; z_3= \frac{p_{4}^2}{M_1^2},\;\; z_4=
\frac{s_{13}}{M_1^2},\;\; z_5= \frac{s_{14}}{M_1^2},\;\; z_6=
\frac{s_{51}}{M_1^2},\;\; z_7= -\frac{s_{24}}{M_1^2},\nonumber\\
&& z_8= -\frac{s_{25}}{M_1^2},\;\; z_9= -\frac{s_{35}}{M_1^2}
,\eeq the superscript $(1)$ means that one external is massive.

Another one, similar to the second we show for the on-shell case,
\beq\label{6off1b} {\cal J}_6^{(1)}(\{\alpha_i\}) &=&
\G_6'(\ai,\avi,j;s_{51}/M_1^2) {\cal T}^{(6-1off)}_9\left[
\begin{array}{l|}-\aiii,-\aiv,-\av,\avi,\avi-\s_6  \\
1+\aii-\s_6, 1+\alpha_{16}-\s_6 \end{array} z_1;...;z_{9}\right],
\eeq where it has as the previous one nine variables, \beq &&
z_1=\frac{p_2^2}{M_1^2},\;\; z_2= \frac{p_{3}^2}{M_1^2},\;\; z_3=
\frac{p_{4}^2}{M_1^2},\;\; z_4= \frac{s_{13}}{M_1^2},\;\; z_5=
\frac{s_{14}}{M_1^2},\;\; z_6= -\frac{s_{24}}{M_1^2},\;\; z_7=
\frac{s_{25}}{s_{51}},\nonumber\\ && z_8=
\frac{s_{35}}{s_{51}},\;\; z_9= \frac{M_1^2}{s_{51}}, \eeq where
$\G_6'(.) = \G_5'(.)$ substituting $\s_5$ by $\s_6$.

\subsection{Six massive external legs}
Instead of presenting the particular cases -- as we have done for
the pentagon -- where two, three, four and five external are
massive, we jump toward the most general case, namely, six massive
external legs,

\be\label{6off6} {\cal J}_6^{(6)}(\{\alpha_i\}) =
\G_6(\ai,\avi;M_5^2)  {\cal S}^{(6-6off)}_{14}\left[\begin{array}{l|}-\aii,-\aiii,-\aiv, -\av, -\s_6 \\
1+\ai-\s_6, 1+\avi-\s_6 \end{array} z_1;...;z_{14}\right], \ee
where \beq && z_1=\frac{p_2^2}{M_5^2},\;\;
z_2=\frac{p_{3}^2}{M_5^2},\;\; z_3=\frac{p_{4}^2}{M_5^2}, \;\;
z_4=-\frac{s_{13}}{M_5^2},\;\; z_5=-\frac{s_{14}}{M_5^2}, \;\;
z_6=\frac{s_{51}}{M_5^2},\;\; z_7=-\frac{s_{24}}{M_5^2},
\nonumber\\ && z_8=\frac{s_{25}}{M_{5}^2},\;\;
z_9=\frac{s_{35}}{M_{5}^2}, \;\;
z_{10}=-\frac{M_{1}^2}{M_5^2},\;\;
z_{11}=-\frac{M_{12}^2}{M_{5}^2},\;\; z_{12}=
-\frac{M_{23}^2}{M_{5}^2},\;\;
z_{13}=\frac{M_{34}^2}{M_{5}^2},\nonumber\\
&& z_{14}=\frac{M_{45}^2}{M_{5}^2}, \eeq where $\Sigma j =
j_{123456789}$ and $\Sigma n=n_{12345}$. From the above result we
 can infer several special cases, namely, on-shell external legs,
 equal masses external legs and so on. To work out these
 particular cases one can proceed on the same way we did in the
 previous sections and subsections.

\subsection{On-shell hexagon with 6 massive propagators}
In this subsection, the last one, we present for completeness a
result for the hexagon scalar integral, where all its external
legs are massless and on-shell and its propagators have distinct
masses, $M_1,...,M_6$.

We pick a sample hypergeometric series, that allows us to obtain
several particular cases (equal masses, or some of them null). The
one that is not contained in this kinematical region is the case
that has $M_6=0$.

Let us call such 14-fold series ${\cal J}_{6m}$


\be \label{6mass} {\cal J}_{6m}(\{\alpha_i\}) =
\G_6^{(mass)}(\avi;M_6) {\cal S}^{(6-6mass)}_{14}\left[
\begin{array}{l|}-\ai,-\aii,-\aiii,-\aiv,-\av,-\s_6 \\
1+\avi-\s_6, D/2 \end{array} z_1;...;z_{14}\right], \ee where \beq
&& z_1=-\frac{p_2^2}{M_6^2} ,\;\; z_2=-\frac{p_{3}^2}{M_6^2} ,\;\;
z_3=-\frac{p_{4}^2}{M_6^2} ,\;\; z_4=-\frac{s_{13}}{M_6^2}
,\;\;z_5=-\frac{s_{14}}{M_6^2} ,\;\;z_6=\frac{s_{51}}{M_6^2}
,\;\;z_7=-\frac{s_{24}}{M_6^2}\nonumber\\
&& z_8=\frac{s_{35}}{M_{6}^2} ,\;\;z_9=\frac{s_{35}}{M_{6}^2}
,\;\;z_{10}=\frac{M_{1}^2}{M_6^2}
,\;\;z_{11}=\frac{M_{2}^2}{M_{6}^2}
,\;\;z_{12}=\frac{M_{3}^2}{M_{6}^2}
,\;\;z_{13}=\frac{M_{4}^2}{M_{6}^2}
,\;\;z_{14}=\frac{M_{5}^2}{M_{6}^2}, \eeq where $\G_6^{(mass)}$ is
given by equation (\ref{gama5mass}) changing $\s_5$ by $\s_6$. So
we finish our study on six-point functions. Special cases where
two or more masses are equal can be obtained summing up the
series, as we did in previous sections.

\section{Conclusions}
Higher one-loop n-point integrals are becoming more important to
scattering process\cite{binoth,bern-penta,bern-fusion} in the
standard model. We presented in this work a general formula for
such integrals, and did show how to apply it to several cases of
interest, from one through six-point integrals. Being its external
legs on or off mass shell, its propagators massless or massive,
and their exponents arbitrary. Working with different masses or
equal ones raises the same difficult in the NDIM approach, as well
as exponents of propagators, taking the most general case does not
imply in additional technical difficulties. We choose sample
hypergeometric solutions, the ones in which one can take
interesting particular cases. Of course, it is not possible to
present all of them (see tables I,II and III). The same procedure
can be followed in order to calculate even higher ($N\geq 7)$
scalar integrals.

\begin{table}[H]
 \caption{Off-shell Scalar Box Integral: number of systems, solutions and kind of series \label{table-box}}
 \begin{ruledtabular}
\begin{tabular}{ccccc}
Off-shell Box & Total & Solutions & No solution & Hypergeom. series\\
  Massless & 252 & 162 & 90 & 5-fold\\
 1 mass & 462 & 267 & 195& 6-fold\\
 2 masses & 792 & 426 & 366 & 7-fold\\
 3 masses & 1287 & 663 & 624& 8-fold\\
 4 masses & 2002 & 1012 & 990& 9-fold\\
 \end{tabular}
 \end{ruledtabular}
 \end{table}

 \begin{table}[H]
 \caption{Pentagon Integral: number of systems, solutions and kind of series \label{table-penta}}
 \begin{ruledtabular}
\begin{tabular}{ccccc}
Pentagon & Total & Solutions & No solution & Hypergeom. series\\
  On-shell & 210 & 125 & 85 & 4-fold\\
 1 off-shell & 462 & 247 & 215& 5-fold\\
 2 off-shell & 924 & 474 & 450& 6-fold\\
 3 off-shell & 1716 & 855 & 861& 7-fold\\
 4 off-shell & 3003 & 1518 & 1485& 8-fold\\
 5 off-shell & 5005 & 2530 & 2475& 9-fold\\
 \end{tabular}
 \end{ruledtabular}
 \end{table}

\begin{table}[H]
 \caption{Hexagon Integral: number of systems, solutions and kind of series \label{table-hex}}
 \begin{ruledtabular}
\begin{tabular}{ccccc}
Hexagon & Total & Solutions & No solution & Hypergeom. series\\
  On-shell & 6435 & 2790 & 3645 & 8-fold\\
 1 off-shell & 11440 & 4736 & 6704 & 9-fold\\
 2 off-shell & 19448 & 7155 & 12293 & 10-fold\\
 3 off-shell & 31824 & 12408 & 19416 & 11-fold\\
 4 off-shell & 50388 & 19484 & 30904& 12-fold\\
 5 off-shell & 77520 & 30410 & 47110 & 13-fold\\
 6 off-shell & 116280 & 45615 & 70665& 14-fold\\
 \end{tabular}
 \end{ruledtabular}
 \end{table}



\begin{acknowledgments}
ATS and AGMS gratefully acknowledge CNPq and ESS wishes to thank
CAPES for financial support.
\end{acknowledgments}

\appendix
\section{Some hypergeometric function used}
We present here some of hypergeometric or hypergeometric-type
functions used in this paper all in terms of the generalized
Lauricella functions \cite{hussain}, that is expressed by
\begin{eqnarray}
&&F^{A:B^{1};...;B^{N}}_{C:D^{1};...;D^{N}} \left[
\begin{array}{l|}
{[a:\alpha^{(1)},...,\alpha^{(N)}];[b^{(1)}:\beta^{(1)}];...;[b^{(N)}:\beta^{(N)}]} \\
{[c:\gamma^{(1)},...,\gamma^{(N)}];[d^{(1)}:\delta^{(1)}];...;[d^{(N)}:\delta^{(N)}]}
\end{array}
z_1;...;z_N\right]\nonumber \\
&&=\sum_{j_1,...,j_{N}=0}^{\infty}\frac{\prod_{i=1}^A(a_i)_{\alpha_{i}^{(1)}j_1+...
+\alpha_{i}^{(N)}j_N}}{\prod_{i=1}^C(c_i)_{\gamma_{i}^{(1)}j_1+...+\gamma_{i}^{(N)}j_N}}
\frac{\prod_{i=1}^{B^1}{(b_i^{1})}_{\beta_{i}^{(1)}j_1}...
{\prod_{i=1}^{B^{(N)}}}{(b_i^{(N)})}_{\beta_{i}^{(N)}j_N}}
{{\prod_{i=1}^{D^{(1)}}}{(d_i^{1})}_{\delta_{i}^{(1)}j_1}...
{\prod_{i=1}^{D^{(N)}}}{(d_i^{(N)})}_{\delta_{i}^{(N)}j_N}}\nonumber \\
&&\times \frac{z_1^{j_1}}{j_{1}!}... \frac{z_{N}^{j_N}}{j_{N}!}
\label{A1}
\end{eqnarray}
where
\begin{eqnarray}
&&[a:\alpha^{(1)},...,\alpha^{(N)}]=(a_1:\alpha_1^{(1)},...,\alpha_1^{(N)}),...,(a_A:\alpha_A^{(1)},...,\alpha_A^{(N)}), \nonumber \\
&&{[b^{(1)}:\beta^{(k)}]}=(b_1^{(k)}:\beta_1^{(k)}),...,(b_{B^{(k)}}^{(k)}:\beta_{B^{(k)}}^{(k)}),\nonumber \\
&&{[c:\gamma^{(1)},...,\gamma^{(N)}]}=(c_1:\gamma_1^{(1)},...,\gamma_1^{(N)}),...,(a_C:\gamma_C^{(1)},...,\gamma_C^{(N)}),\nonumber \\
&&{[d^{(1)}:\delta^{(k)}]}=(d_1^{(k)}:\delta_1^{(k)}),...,(d_{D^{(k)}}^{(k)}:\delta_{D^{(k)}}^{(k)}),\nonumber
\end{eqnarray}
with $k=1,...,N $. The parameters $\alpha, \beta, \gamma, \delta$
can be non-negative integers in hypergeometric functions and too
negative integers in hypergeometric-type functions. From
(\ref{A1}) we can extract all the functions used in this paper,
that is,
\begin{eqnarray}
R_{6}&=&R_{6}\left[
\begin{array}{l|}
x_1,x_2,x_3 \\
x_4,x_5
\end{array}
z_1;z_2;z_3;z_4\right] \nonumber \\
&=&F^{4:0;0;0;0}_{1;0;0;0;0}\left[
\begin{array}{l|}
(x_{1}:1,1,1,1),(x_2:1,1,1,0),(x_3:1,0,0,1) \\
(x_4:1,0,1,1),(x_5:1,1,0,0)
\end{array}
z_1;z_2;z_3;z_4\right], \nonumber \\
\end{eqnarray}
\begin{eqnarray}
R_{11}&=&R_{11}\left[
\begin{array}{l|}
x_1,x_2,x_3,x_4 \\
x_5,x_6,x_7
\end{array}
z_1;...;z_4\right] \nonumber \\
&=&F^{4:0}_{3;0}\left[
\begin{array}{l|}
(x_{1}:1),(x_2:1,1,0,1),(x_3:1,0,1,1),(x_4:1,0,1,0) \\
(x_5:1,0,1,1),(x_6:1,0,2,1),(x_7:0,1,-1,0)
\end{array}
z_1;...;z_4\right], \nonumber \\
\end{eqnarray}
where $F^{4:0}_{3;0}=F^{4:0;0;0;0}_{3;0;0;0;0}$ and
$(x_{1}:1)=(x_{1}:1,1,1,1)$,
\begin{eqnarray}
\Psi_{3}&=&\Psi_{3}\left[
\begin{array}{l|}
x_1,x_2,x_3 \\
x_4,x_5
\end{array}
z_1;...;z_5\right] \nonumber \\
&=&F^{3:0}_{2;0}\left[
\begin{array}{l|}
(x_{1}:1,1,1,1,1),(x_2:1,1,0,1,0),(x_3:1,0,1,0,1) \\
(x_4:1,0,0,1,1),(x_5:1,1,1,0,0)
\end{array}
z_1;...;z_5\right], \nonumber \\
\end{eqnarray}
where $F^{3:0}_{2;0}=F^{3:0;0;0;0}_{2;0;0;0;0}$,
\begin{eqnarray}
\Phi&=&\Phi\left[
\begin{array}{l|}
x_1,x_2,x_3,x_4 \\
x_5,x_6
\end{array}
z_1;...;z_{9}\right] \nonumber \\
&=&F^{4:0}_{2;0}\left[
\begin{array}{l|}
(x_{1}:a_1),(x_2:a_2),(x_3:a_3)(x_4:a_4) \\
(x_5:a_5),(x_6:a_6)
\end{array}
z_1;...;z_9\right], \nonumber \\
\end{eqnarray}
where
$F^{4:0;...;0}_{2;0;...;0}=F^{4:0;0;0;0;0;0;0;0;0}_{2;0;0;0;0;0;0;0;0;0}$
and $(x_{1}:a_1)=(x_{1}:1,1,1,1,1,1,1,1,1)$,
$(x_2:a_2)=(x_2:1,1,0,1,0,0,0,1,0,0)$,
$(x_3:a-3)=(x_3:1,0,0,1,1,0,0,1,0)$,
$(x_4:a_4)=(x_4:0,1,0,1,0,1,0,0,1)$,
$(x_5:a_5)=(x_5:1,1,1,1,1,1,0,0,0)$ and
$(x_6:a_6)=(x_6:1,1,0,1,0,0,0,0,1)$.

For the pentagon integrals we define the hypergeometric series,
${{\cal S}_A^{(5)}}$, where "5" and $A$ represent pentagon and the
number of variables the series has, respectively.

In equation (\ref{penta-on1}) the result was written in terms of,
\begin{eqnarray}
\!\!\!\!\!\!\!\! {\cal S}^{(5)}_4 &=& {\cal S}^{(5)}_4\left[
\begin{array}{l|}x_1,x_2,x_3,x_4 \\
x_5,x_6\end{array}
z_1;...;z_{4}\right] \nonumber \\
&=&F^{4:0}_{2;0}\left[
\begin{array}{l|}
(x_{1}:1,1,0,0),(x_2:0,0,1,1),(x_3:0,1,0,1),(x_4:1) \\
(x_5:0,1,1,1),(x_6:1,1,0,1)\end{array}
z_1;...;z_4\right], \nonumber \\
\end{eqnarray} 

Equation (\ref{penta-on2}) was written as a series defined by,
\begin{eqnarray}
\!\!\!\!\!\!\!\! {\cal T}^{(5)}_4 &=& {\cal T}^{(5)}_4\left[
\begin{array}{l|}x_1,x_2,x_3,x_4 \\
x_5,x_6\end{array}
z_1;...;z_{4}\right] \nonumber \\
&=&F^{4:0}_{2;0}\left[
\begin{array}{l|}
(x_{1}:1,1,0,0),(x_2:1,0,1,0),(x_3:0,1,0,1),(x_4:-1,0,0,1) \\
(x_5:0,0,1,1),(x_6:0,1,-1,0)\end{array}
z_1;...;z_4\right], \nonumber \\
\end{eqnarray} 

Five variables series was obtained as a result for the massive
pentagon in the special case where all five masses are equal,
\begin{eqnarray}
\!\!\!\!\!\!\!\! {\cal S}^{(5)}_5 &=& {\cal S}^{(5)}_5\left[
\begin{array}{l|}x_1,x_2,x_3,x_4,x_5,x_6 \\
x_7\end{array}
z_1;...;z_{5}\right] \nonumber \\
&=&F^{6:0}_{1;0}\left[
\begin{array}{l|}
(x_{1}:11000),(x_2:00110),(x_3:10001),(x_4:01010),(x_5:00101),(x_6:1) \\
(x_7:2)\end{array}
z_1;...;z_5\right], \nonumber \\
\end{eqnarray}
where we did not write the commas, i.e., $(x_4:01010) =
(x_4:0,1,0,1,0)$ and so forth. 

Six variables is one of the representation for the pentagon with
two distinct masses,
\begin{eqnarray}
\!\!\!\!\!\!\!\! {\cal S}^{(5)}_6 &=& {\cal S}^{(5)}_6\left[
\begin{array}{l|}x_1,x_2,x_3,x_4,x_5,x_6 \\
x_7,x_8,x_9\end{array}
z_1;...;z_{6}\right] \nonumber \\
&=&F^{6:0}_{3;0}\left[
\begin{array}{l|}
(x_{1}:110000),(x_2:001100),(x_3:100010),(x_4:010101),(x_5:001010),(x_6:1) \\
(x_7:111110),(x_8:212120),(x_9:-1,0,-1,0,-1,1)\end{array}
z_1;...;z_6\right], \nonumber \\
\end{eqnarray} 

The reader can observe that the following result and previous two
follow a pattern, seven variables, for three different masses,
\begin{eqnarray}
\!\!\!\!\!\!\!\! {\cal S}^{(5)}_7 &=& {\cal S}^{(5)}_7\left[
\begin{array}{l|}x_1,x_2,x_3,x_4,x_5,x_6 \\
x_7,x_8,x_9\end{array}
z_1;...;z_{7}\right] \nonumber \\
&=&F^{6:0}_{3;0}\left[
\begin{array}{l|}
(x_{1}:1100000),(x_2:0011000),(x_3:1000110),(x_4:0101001),(x_5:0010100),(x_6:1) \\
(x_7:1111100),(x_8:1121100),(x_9:00,-1,0011)\end{array}
z_1;...;z_7\right], \nonumber \\
\end{eqnarray} 

The eight-fold series representing the case where the pentagon has
four different masses,
\begin{eqnarray}
\!\!\!\!\!\!\!\! {\cal S}^{(5)}_8 &=& {\cal S}^{(5)}_8\left[
\begin{array}{l|}(a_{1}),(a_2),(a_3),(a_4),(a_5),(a_6) \\
(a_7),(a_8),(a_9)\end{array}
z_1;...;z_{8}\right] \nonumber \\
&=&F^{6:0}_{3;0}\left[
\begin{array}{l|}
(a_{1}),(a_2),(a_3),(a_4),(a_5),(a_6) \\
(x_7:11111000),(x_8:1111100),(x_9:00010111)\end{array}
z_1;...;z_8\right], \nonumber \\
\end{eqnarray}
where, $a_1= (x_{1}:11000000),\; a_2=(x_2:00110100), \;
a_3=(x_3:10001010),\; a_4=(x_4:01010001), a_5=(x_5:00101000),\;
a_6=(x_6:1)$. 

Finally the nine-fold series, representing the on-shell pentagon
with five distinct masses,
\begin{eqnarray}
\!\!\!\!\!\!\!\! {\cal S}^{(5)}_9 &=& {\cal S}^{(5)}_9\left[
\begin{array}{l|}(b_{1}),(b_2),(b_3),(b_4),(b_5) \\
(x_7),(x_8)\end{array}
z_1;...;z_{9}\right] \nonumber \\
&=&F^{5:0}_{2;0}\left[
\begin{array}{l|}
(b_{1}),(b_2),(b_3),(b_4),(b_5) \\
(x_7:111110000),(x_8:1101011111)\end{array}
z_1;...;z_9\right], \nonumber \\
\end{eqnarray}
where, $b_1= (x_{1}:110001000),\; b_2=(x_2:001100100), \;
b_3=(x_3:100010010),\; b_4=(x_4:010100001), b_5=(x_5:1)$. 

The off-shell pentagon starts with a five variables Lauricella
function, in the case where one leg is off mass shell,
\begin{eqnarray}
\!\!\!\!\!\!\!\! {\cal S}^{(5-1off)}_{5} &=& {\cal
S}^{(5-1off)}_5\left[
\begin{array}{l|}x_1,x_2,x_3,x_4 \\
x_5,x_6\end{array}
z_1;...;z_{5}\right] \nonumber \\
&=&F^{4:0}_{2;0}\left[
\begin{array}{l|}
(x_{1}:10001),(x_2:01100),(x_3:00011),(x_4:1) \\
(x_5:01110),(x_6:11001)\end{array}
z_1;...;z_5\right], \nonumber \\
\end{eqnarray} 
the following function represents eq.(\ref{1-off-penta}),
\begin{eqnarray}
\!\!\!\!\!\!\!\! {\cal T}^{(5-1off)}_{5} &=& {\cal
T}^{(5-1off)}_5\left[
\begin{array}{l|}x_1,x_2,x_3,x_4 \\
x_5,x_6\end{array}
z_1;...;z_{5}\right] \nonumber \\
&=&F^{4:0}_{2;0}\left[
\begin{array}{l|}
(x_{1}:11001),(x_2:10010),(x_3:00110),(x_4:1) \\
(x_5:11010),(x_6:10111)\end{array}
z_1;...;z_5\right], \nonumber \\
\end{eqnarray} 
we presented two results when two external legs are off-shell, the
first one was given in terms of,
\begin{eqnarray}
\!\!\!\!\!\!\!\! {\cal S}^{(5-2off)}_{6} &=& {\cal
S}^{(5-2off)}_6\left[
\begin{array}{l|}x_1,x_2,x_3,x_4 \\
x_5,x_6\end{array}
z_1;...;z_{6}\right] \nonumber \\
&=&F^{4:0}_{2;0}\left[
\begin{array}{l|}
(x_{1}:100011),(x_2:011000),(x_3:000110),(x_4:1) \\
(x_5:001111),(x_6:110010)\end{array}
z_1;...;z_6\right], \nonumber \\
\end{eqnarray} 
and the second one,
\begin{eqnarray}
\!\!\!\!\!\!\!\! {\cal T}^{(5-1off)}_{6} &=& {\cal
T}^{(5-1off)}_6\left[
\begin{array}{l|}a_1,a_2,a_3,a_4,a_5 \\
x_6\end{array}
z_1;...;z_{6}\right] \nonumber \\
&=&F^{5:0}_{1;0}\left[
\begin{array}{l|}
(a_1),(a_2),(a_3),(a_4),(a_5) \\
(x_6:111010)\end{array}
z_1;...;z_6\right], \nonumber \\
\end{eqnarray}
where $(a_1)=(x_{1}:10010,-1),\;\;
(a_2)=(x_2:010011),\;\;(a_3)=(x_3:110000),\;\;(a_4)=(x_4:001100),\;\;(a_5)=
(x_5:000,-1,11),\;\; (a_6)= (x_6:111010)$. 

The special case of $M=M_{12}$ gives rise to,
\begin{eqnarray} \!\!\!\!\!\!\!\! {\cal U}^{(5-2off)}_{5} &=& {\cal
U}^{(5-2off)}_5\left[
\begin{array}{l|}x_1,x_2,x_3,x_4,x_5 \\
x_6,x_7\end{array}
z_1;...;z_{5}\right] \nonumber \\
&=&F^{5:0}_{2;0}\left[
\begin{array}{l|}
(x_{1}:1001,-1),(x_2:01101),(x_3:11000),(x_4:00110),(x_5:000,-1,1) \\
(x_6:1000,-1),(x_7:1111,-1)\end{array}
z_1;...;z_5\right]. \nonumber \\
\end{eqnarray} 

Three external legs off shell was written as a function,
\begin{eqnarray} \!\!\!\!\!\!\!\! {\cal S}^{(5-3off)}_{7} &=& {\cal
S}^{(5-3off)}_7\left[
\begin{array}{l|}x_1,x_2,x_3,x_4, \\
x_5,x_6\end{array}
z_1;...;z_{7}\right] \nonumber \\
&=&F^{4:0}_{2;0}\left[
\begin{array}{l|}
(x_{1}:1100010),(x_2:0011011),(x_3:00011000),(x_4:1) \\
(x_5:0111010),(x_6:1001111)\end{array}
z_1;...;z_7\right]. \nonumber \\
\end{eqnarray} 

Four external legs, the first result presented,
\begin{eqnarray}
\!\!\!\!\!\!\!\! {\cal S}^{(5-4off)}_{8} &=& {\cal
S}^{(5-4off)}_8\left[
\begin{array}{l|}a_1,a_2,a_3,a_4 \\
x_5,x_6 \end{array}
z_1;...;z_{8}\right] \nonumber \\
&=&F^{4:0}_{2;0}\left[
\begin{array}{l|}
(a_1),(a_2),(a_3),(a_4) \\
(x_5:00111111), (x_6:11001011)\end{array}
z_1;...;z_8\right], \nonumber \\
\end{eqnarray} 
where $(a_1)=(x_{1}:10001110),\;\;
(a_2)=(x_2:01100011),\;\;(a_3)=(x_3:00011001),\;\;(a_4)=(x_4:1)$,
and the second is written as the generalized Lauricella function,
\begin{eqnarray}
\!\!\!\!\!\!\!\! {\cal T}^{(5-4off)}_{8} &=& {\cal
T}^{(5-4off)}_8\left[
\begin{array}{l|}a_1,a_2,a_3,a_4 \\
x_5,x_6 \end{array}
z_1;...;z_{8}\right] \nonumber \\
&=&F^{4:0}_{2;0}\left[
\begin{array}{l|}
(a_1),(a_2),(a_3),(a_4) \\
(x_5:00111001), (x_6:11101110)\end{array}
z_1;...;z_8\right], \nonumber \\
\end{eqnarray} 
where $(a_1)=(x_{1}:11001000),\;\; (a_2)=(x_2:00111100),
\;\;(a_3)=(x_3:01100011),\;\;(a_4)=(x_4:1)$.

Finally, we define the hypergeometric functions for the five
external legs off-shell pentagon,
\begin{eqnarray}
\!\!\!\!\!\!\!\! {\cal S}^{(5-5off)}_{9} &=& {\cal
S}^{(5-5off)}_9\left[
\begin{array}{l|}a_1,a_2,a_3,a_4 \\
x_5,x_6 \end{array}
z_1;...;z_{9}\right] \nonumber \\
&=&F^{4:0}_{2;0}\left[
\begin{array}{l|}
(a_1),(a_2),(a_3),(a_4) \\
(x_5:001110111), (x_6:111001110)\end{array}
z_1;...;z_9\right], \nonumber \\
\end{eqnarray} 
where $(a_1)=(x_{1}:001101100),\;\; (a_2)=(x_2:100010110),
\;\;(a_3)=(x_3:011000011),\;\;(a_4)=(x_4:1)$, the second one,
\begin{eqnarray}
\!\!\!\!\!\!\!\! {\cal T}^{(5-5off)}_{9} &=& {\cal
T}^{(5-5off)}_9\left[
\begin{array}{l|}a_1,a_2,a_3,a_4 \\
x_5,x_6 \end{array}
z_1;...;z_{9}\right] \nonumber \\
&=&F^{4:0}_{2;0}\left[
\begin{array}{l|}
(a_1),(a_2),(a_3),(a_4) \\
(x_5:0,-1,011010,-1), (x_6:111001110)\end{array}
z_1;...;z_9\right], \nonumber \\
\end{eqnarray} 
where $(a_1)=(x_{1}:110001001),\;\; (a_2)=(x_2:001101100),
\;\;(a_3)=(x_3:100010110),\;\;(a_4)=(x_4:00111011,-1)$, and the
third,
\begin{eqnarray}
\!\!\!\!\!\!\!\! {\cal U}^{(5-5off)}_{9} &=& {\cal
U}^{(5-5off)}_9\left[
\begin{array}{l|}a_1,a_2,a_3,a_4,a_5 \\
x_6,x_7 \end{array}
z_1;...;z_{9}\right] \nonumber \\
&=&F^{5:0}_{2;0}\left[
\begin{array}{l|}
(a_1),(a_2),(a_3),(a_4) \\
(x_6:111100110), (x_7:0100110,-1,-1)\end{array}
z_1;...;z_9\right], \nonumber \\
\end{eqnarray} 
where $(a_1)=(x_{1}:01101110,-1),\;\; (a_2)=(x_2:100100011),
\;\;(a_3)=(x_3:011110000),\;\;(a_4)=(x_4:110001100)$

The hexagon integrals are also defined in terms of generalized
Lauricella functions, the firs result we presented for the
on-shell case is given as,
\begin{eqnarray}
\!\!\!\!\!\!\!\! {\cal S}^{(6-on)}_{8} &=& {\cal
S}^{(6-on)}_8\left[
\begin{array}{l|}a_1,a_2,a_3,a_4,a_5 \\
x_6,x_7 \end{array}
z_1;...;z_{8}\right] \nonumber \\
&=&F^{5:0}_{2;0}\left[
\begin{array}{l|}
(a_1),(a_2),(a_3),(a_4),(a_5) \\
(x_6:10101111), (x_7:11111010)\end{array}
z_1;...;z_8\right], \nonumber \\
\end{eqnarray} 
where $(a_1)=(x_{1}:11100000),\;\; (a_2)=(x_2:00011100),
\;\;(a_3)=(x_3:10000011),\;\;(a_4)=(x_4:00101010), (a_5)=(x_5:1)$.
The second one,
\begin{eqnarray}
\!\!\!\!\!\!\!\! {\cal T}^{(6-on)}_{8} &=& {\cal
T}^{(6-on)}_8\left[
\begin{array}{l|}a_1,a_2,a_3,a_4,a_5 \\
x_6,x_7 \end{array}
z_1;...;z_{8}\right] \nonumber \\
&=&F^{5:0}_{2;0}\left[
\begin{array}{l|}
(a_1),(a_2),(a_3),(a_4),(a_5) \\
(x_6:01111110), (x_7:1101000,-1)\end{array}
z_1;...;z_8\right], \nonumber \\
\end{eqnarray}
where $(a_1)=(x_{1}:1101011,-1),\;\; (a_2)=(x_2:11100000),
\;\;(a_3)=(x_3:00011100),\;\;(a_4)=(x_4:01010010),
(a_5)=(x_5:00101001)$.

When one of the external particles is massive, or an external leg
is off mass shell, the integral is written in terms of 9-fold
Lauricella function,
\begin{eqnarray}
\!\!\!\!\!\!\!\! {\cal S}^{(6-1off)}_{9} &=& {\cal
S}^{(6-1off)}_9\left[
\begin{array}{l|}a_1,a_2,a_3,a_4,a_5 \\
x_6,x_7 \end{array}
z_1;...;z_{9}\right] \nonumber \\
&=&F^{5:0}_{2;0}\left[
\begin{array}{l|}
(a_1),(a_2),(a_3),(a_4),(a_5) \\
(x_6:00011111), (x_7:111000111)\end{array}
z_1;...;z_9\right], \nonumber \\
\end{eqnarray}
where $(a_1)=(x_{1}:100000110),\;\; (a_2)=(x_2:010100001),
\;\;(a_3)=(x_3:001010100),\;\;(a_4)=(x_4:000001011),
(a_5)=(x_5:1)$, the next we presented is also for the case of one
massive external particle,
\begin{eqnarray}
\!\!\!\!\!\!\!\! {\cal T}^{(6-1off)}_{9} &=& {\cal
T}^{(6-1off)}_9\left[
\begin{array}{l|}a_1,a_2,a_3,a_4,a_5 \\
x_6,x_7 \end{array}
z_1;...;z_{9}\right] \nonumber \\
&=&F^{5:0}_{2;0}\left[
\begin{array}{l|}
(a_1),(a_2),(a_3),(a_4),(a_5) \\
(x_6:111001110), (x_7:00011100,-1)\end{array}
z_1;...;z_9\right], \nonumber \\
\end{eqnarray}
where $(a_1)=(x_{1}:100001100),\;\; (a_2)=(x_2:010100010),
\;\;(a_3)=(x_3:001011000),\;\;(a_4)=(x_4:000000111),
(a_5)=(x_5:11111100,-1)$.

The most complicated integral, the hexagon with all external legs
massive was given as a 14-fold series,
\begin{eqnarray}
\!\!\!\!\!\!\!\! {\cal S}^{(6-6off)}_{14} &=& {\cal
S}^{(6-6off)}_{14}\left[
\begin{array}{l|}a_1,a_2,a_3,a_4,a_5 \\
x_6,x_7 \end{array}
z_1;...;z_{14}\right] \nonumber \\
&=&F^{5:0}_{2;0}\left[
\begin{array}{l|}
(a_1),(a_2),(a_3),(a_4),(a_5) \\
(x_6:00011111101111), (x_7:11111010011110)\end{array}
z_1;...;z_{14}\right], \nonumber \\
\end{eqnarray}
where $(a_1)=(x_{1}:00011111000),\;\; (a_2)=(x_2:10000011001100),
\;\;(a_3)=(x_3:01010000100110),\;\;(a_4)=(x_4:00101010000011),
(a_5)=(x_5:1)$.

Finally, the last generalized Lauricella function, for on-shell
hexagon with six different massive propagators,
\begin{eqnarray}
\!\!\!\!\!\!\!\! {\cal S}^{(6-6mass)}_{14} &=& {\cal
S}^{(6-6mass)}_{14}\left[
\begin{array}{l|}a_1,a_2,a_3,a_4,a_5,a_6 \\
x_6,x_7 \end{array}
z_1;...;z_{14}\right] \nonumber \\
&=&F^{6:0}_{2;0}\left[
\begin{array}{l|}
(a_1),(a_2),(a_3),(a_4),(a_5),(a_6) \\
(x_7:11111110011111), (x_8:11111111100000)\end{array}
z_1;...;z_{14}\right], \nonumber \\
\end{eqnarray}
where $(a_1)=(x_{1}:11100000010000),\;\;
(a_2)=(x_2:00011100001000), \;\;(a_3)=(x_3:10000011000100),\;\;
(a_4)=(x_4:01010000100010),\;\; (a_5)=(x_5:00101010000001),\;\;
(a_6)=(x_6:1) $.

\bibliography{penta-hex.bbl}

\end{document}